\documentclass[
  aps,prx,reprint,superscriptaddress,footinbib,
  longbibliography 
]{revtex4-2}

\usepackage{graphicx}
\usepackage{amsmath,amssymb,bm,mathtools}
\usepackage{xcolor}
\usepackage{tikz}
\usetikzlibrary{arrows.meta,positioning,calc}
\definecolor{innergreen}{RGB}{225,246,218}
\definecolor{edgegreen}{RGB}{0,116,0}
\definecolor{outerpink}{RGB}{252,232,228}
\definecolor{edgered}{RGB}{179,32,8}
\definecolor{outerblue}{RGB}{215,223,245}
\definecolor{edgeblue}{RGB}{0,80,129}
\usepackage{float}
\usepackage{hyperref}
\hypersetup{colorlinks=true,linkcolor=blue,citecolor=blue,urlcolor=blue}

\begin{document}

\title{From Fermions to Vortices: Microscopic Simulations of Phase Ordering \\ in a \(d\)-Wave Superconductor}

\author{Ali Rayat}
\affiliation{Department of Physics, University of Virginia, Charlottesville, Virginia 22904, USA}

\author{Gia-Wei Chern}
\affiliation{Department of Physics, University of Virginia, Charlottesville, Virginia 22904, USA}

\date{\today}

\begin{abstract}
Describing bath-induced superconducting phase ordering directly from a microscopic fermionic Hamiltonian remains a formidable challenge. Here we develop a microscopic framework for this problem using the recently introduced adiabatic variational principle (AVP) and demonstrate its application to a two-dimensional $d$-wave superconductor. The central construction maps evolving, spatially inhomogeneous bond-pairing fields onto constrained Bogoliubov--de Gennes states, with the associated Lagrange multipliers determining the thermodynamic driving forces. This enables dissipative evolution of the superconducting order while retaining its microscopic electronic structure, without assuming a phenomenological Ginzburg--Landau free-energy functional. Starting from disordered pairing configurations, we resolve the emergence of phase coherence through vortex--antivortex annihilation and the accompanying evolution of local electronic spectra. Vortex decay and dynamical scaling recover the characteristic kinetics of two-dimensional phase ordering, providing a nontrivial benchmark of the framework. The solver-independent AVP formulation opens a route to microscopic studies of dissipative ordering and metastability in superconductors with intertwined stripe and pair-density-wave orders.
\end{abstract}

\maketitle

Superconductivity is a prominent manifestation of collective quantum order and spontaneous symmetry breaking \cite{Bardeen1957,deGennes1966}. Its emergence requires not only electron pairing but also spatial phase coherence of a complex order parameter \cite{Emery1995}. Following rapid cooling into the superconducting regime, an initially disordered pairing field can develop coherence through vortex--antivortex annihilation and the growth of phase-coherent regions \cite{Bray1994,Yurke1993,Jelic2011}. This process is particularly intriguing in unconventional $d$-wave superconductors, whose sign-changing pairing can be represented by complex fields on lattice bonds \cite{Sigrist1991,Tsuei2000,Scalapino2012,Keimer2015}. The evolving textures reshape the electronic spectrum, including the low-energy quasiparticle states associated with nodal pairing and vortex cores \cite{Wang1995,Franz1998,Pan2000}. Understanding the joint development of superconducting coherence and electronic structure thus presents an important nonequilibrium many-body problem.

Superconducting phase ordering is commonly described using time-dependent Ginzburg--Landau (TDGL) theories and related phase-field approaches \cite{Schmid1966,Hohenberg1977,Tinkham1996}. These theories capture vortex motion, defect annihilation, and dynamical scaling through a coarse-grained free-energy functional and a phenomenological evolution equation for the order parameter. Microscopic derivations rely on restrictive expansions in the pairing amplitude and its spatial and temporal variations \cite{Gorkov1959,Abrahams1966}, which generally break down for strongly inhomogeneous textures during relaxation. Such descriptions absorb the electronic degrees of freedom into effective coefficients rather than explicitly resolving them. Resolving the electronic structure of evolving superconducting textures therefore calls for a direct connection to the underlying fermionic Hamiltonian.

Interacting lattice models provide a natural setting for studying unconventional pairing and its competition with charge and magnetic order. Within a mean-field treatment, Bogoliubov--de Gennes (BdG) theory resolves spatially nonuniform pairing configurations and their electronic quasiparticle states \cite{Micnas1990,Dagotto1994,Dagotto1994b,Bishop2016,deGennes1966,Gygi1991,Ghosal2000,Berthod2016}. Self-consistent solutions describe equilibrium or metastable stationary states, which may themselves be spatially inhomogeneous. Its dynamical extension, time-dependent BdG theory, can be formulated within the general time-dependent variational principle (TDVP) and has been widely applied to superconducting dynamics following interaction quenches \cite{Barankov2004,Barankov2006,Yuzbashyan2006,Foster2013,Foster2014,Chern2019}. These coherent, closed-system dynamics do not describe the dissipative relaxation considered here. The central task is to construct a thermodynamic landscape over evolving order parameters, with microscopic electronic states that consistently reproduce the instantaneous order and supply the forces driving its relaxation.

The recently developed adiabatic variational principle (AVP) provides a solver-independent construction of this landscape by imposing the order parameters as constraints on the microscopic many-body state \cite{Yang2026b}. Assuming rapid electronic relaxation relative to the collective evolution, AVP associates each prescribed configuration with a constrained quasi-equilibrium state. Its pairing expectation values reproduce the instantaneous order, while the conjugate Lagrange multipliers determine the free-energy derivatives. The resulting microscopic forces require neither a phenomenological Ginzburg--Landau functional nor a local gradient expansion. Nonlinear Mori--Zwanzig projection theory provides a systematic basis for the associated reduced dynamics, including reversible evolution, dissipation, and fluctuations \cite{Zwanzig1960,Mori1965,Grabert1978,Yang2026b}.

Here we develop a constrained BdG implementation of this framework to investigate phase ordering in the square-lattice extended Hubbard model with attractive nearest-neighbor interactions. We retain independent complex bond-pairing fields, allowing local departures from a pure $d$-wave configuration, and evolve them through purely relaxational, Model-A dynamics. Starting from disordered configurations, the simulations resolve the growth of superconducting phase coherence through vortex--antivortex annihilation, together with the evolution of local electronic spectra. Vortex decay and correlation scaling provide nontrivial benchmarks against the established kinetics of two-dimensional phase ordering \cite{Bray1994,Yurke1993,Jelic2011}. These results demonstrate a microscopic route from fermionic pairing to collective vortex dynamics, with the evolving order and electronic state kept consistent throughout the relaxation.

\begin{figure}[t]
\centering
\includegraphics[width=\columnwidth]{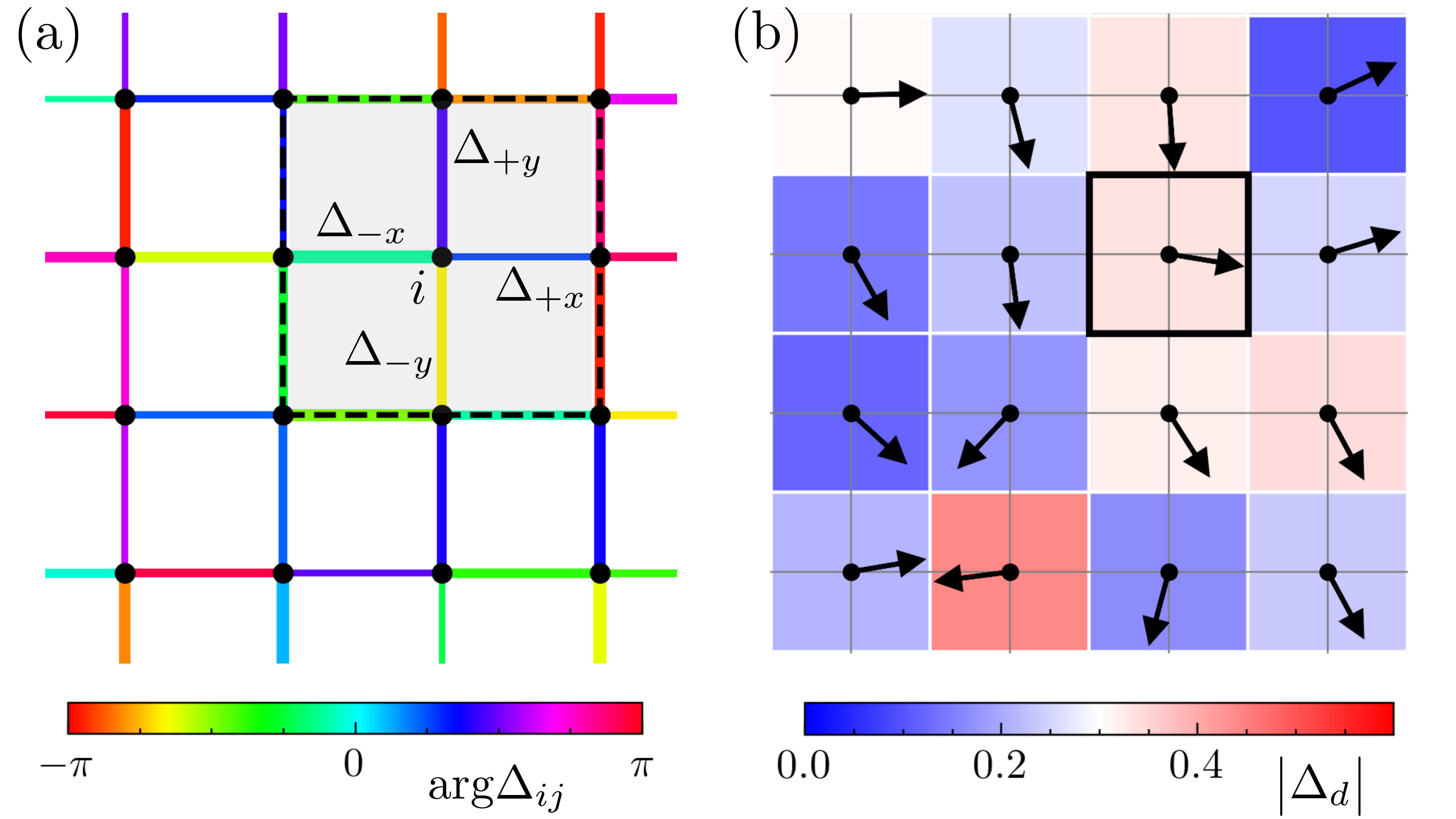}
\caption{Mapping from bond pairing to local $d$-wave order. (a) Bond fields $\Delta_{ij}$, colored by phase. The four bonds adjacent to site $i$, labeled $\Delta_{\pm x}$ and $\Delta_{\pm y}$, determine $\Delta_i^d$ through Eq.~\eqref{eq:SC_local_d}. (b) Site-centered $d$-wave field, with color indicating its magnitude and arrows its phase. The black square identifies site $i$.}
\label{fig:SC_mapping}
\end{figure}

We consider the square-lattice extended Hubbard model with attractive nearest-neighbor interactions and vanishing on-site interaction \cite{Micnas1988,Micnas1990,Chen2023},
\begin{equation}
\hat{\mathcal H}=-t\sum_{\langle ij\rangle,\sigma}(\hat c_{i\sigma}^\dagger\hat c_{j\sigma}+{\rm h.c.})-V\sum_{\langle ij\rangle}\hat n_i\hat n_j-\mu\sum_i\hat n_i,
\label{eq:SC_model}
\end{equation}
where $\hat n_i=\sum_\sigma\hat c_{i\sigma}^\dagger\hat c_{i\sigma}$, $V>0$, and each undirected bond is counted once. The density attraction supports both singlet and triplet bond-pairing channels. Among the competing pairing states, singlet $d_{x^2-y^2}$ order is favored near half filling \cite{Micnas1988,Micnas1990,Huang2013}. We therefore focus on the spin-singlet channel \footnote{See Appendix~\ref{app:SC_pairing_channels} for the singlet--triplet decomposition and comparisons of pairing susceptibilities and uniform mean-field free energies.}, with bond operator and order parameter
\begin{equation}
\hat B_{ij}=\hat c_{j\downarrow}\hat c_{i\uparrow}-\hat c_{j\uparrow}\hat c_{i\downarrow},\qquad \Delta_{ij}=-\frac V2\langle\hat B_{ij}\rangle.
\label{eq:SC_bond_order}
\end{equation}
The symmetric complex fields $\Delta_{ij}=\Delta_{ji}$ are the dynamical variables. A uniform $d_{x^2-y^2}$ state has opposite pairing signs on horizontal and vertical bonds. To characterize spatially inhomogeneous order, we use the local projection
\begin{equation}
\Delta_i^d=\frac14\left(\Delta_{i,i+\hat x}+\Delta_{i,i-\hat x}-\Delta_{i,i+\hat y}-\Delta_{i,i-\hat y}\right),
\label{eq:SC_local_d}
\end{equation}
illustrated in Fig.~\ref{fig:SC_mapping}. This projection extracts the local $d$-wave component of a general bond configuration, which may also contain an extended-$s$ component.

We assume that the collective pairing fields evolve slowly compared with the relaxation of the remaining electronic degrees of freedom. This separation of time scales motivates an adiabatic description, analogous to the Born--Oppenheimer approximation: the quasiparticles equilibrate at each instantaneous bond configuration, providing the thermodynamic landscape for its slow evolution. Because the pairing fields are themselves electronic expectation values, the quasi-equilibrium state must reproduce the prescribed $\Delta_{ij}$ through Eq.~\eqref{eq:SC_bond_order}. Enforcing this consistency by constrained variational minimization is the central construction of the recently introduced adiabatic variational principle (AVP) \cite{Yang2026b}.

At fixed temperature and chemical potential, the constrained minimum of the electronic grand potential defines $F[\Delta]$. Its derivatives supply the thermodynamic forces driving the collective relaxation. We implement this construction within the pairing-channel mean-field approximation by restricting the electronic state to a normalized Gaussian density operator $\hat\rho_{\rm G}$. The auxiliary Lagrangian is
\begin{equation}
\begin{aligned}
\mathcal L&=\operatorname{Tr}(\hat\rho_{\rm G}\hat{\mathcal H})+T\operatorname{Tr}(\hat\rho_{\rm G}\ln\hat\rho_{\rm G})\\
&\quad-\sum_{\langle ij\rangle}\left[f_{ij}^*\left(\Delta_{ij}+\frac V2\langle\hat B_{ij}\rangle\right)+{\rm c.c.}\right],
\end{aligned}
\label{eq:SC_constrained_functional}
\end{equation}
where the interaction expectation is evaluated in the retained spin-singlet pairing channel. The complex Lagrange multipliers $f_{ij}$ enforce both components of each prescribed bond order and are the thermodynamic forces conjugate to $\Delta_{ij}$. Variation within the Gaussian manifold yields a thermal state of the constrained BdG Hamiltonian,
\begin{equation}
\begin{aligned}
\hat{\mathcal H}_{\rm BdG}&=-t\sum_{\langle ij\rangle,\sigma}(\hat c_{i\sigma}^\dagger\hat c_{j\sigma}+{\rm h.c.})-\mu\sum_i\hat n_i\\
&\quad+\sum_{\langle ij\rangle}(\Delta_{ij}^{\rm eff}\hat B_{ij}^\dagger+{\rm h.c.}),
\end{aligned}
\label{eq:SC_BdG_Hamiltonian}
\end{equation}
with $\Delta_{ij}^{\rm eff}=-(V/2)(\langle\hat B_{ij}\rangle+f_{ij})$. The normal-state band is kept fixed in this pairing-channel approximation.

\begin{figure*}[t]
\centering
\includegraphics[width=\textwidth]{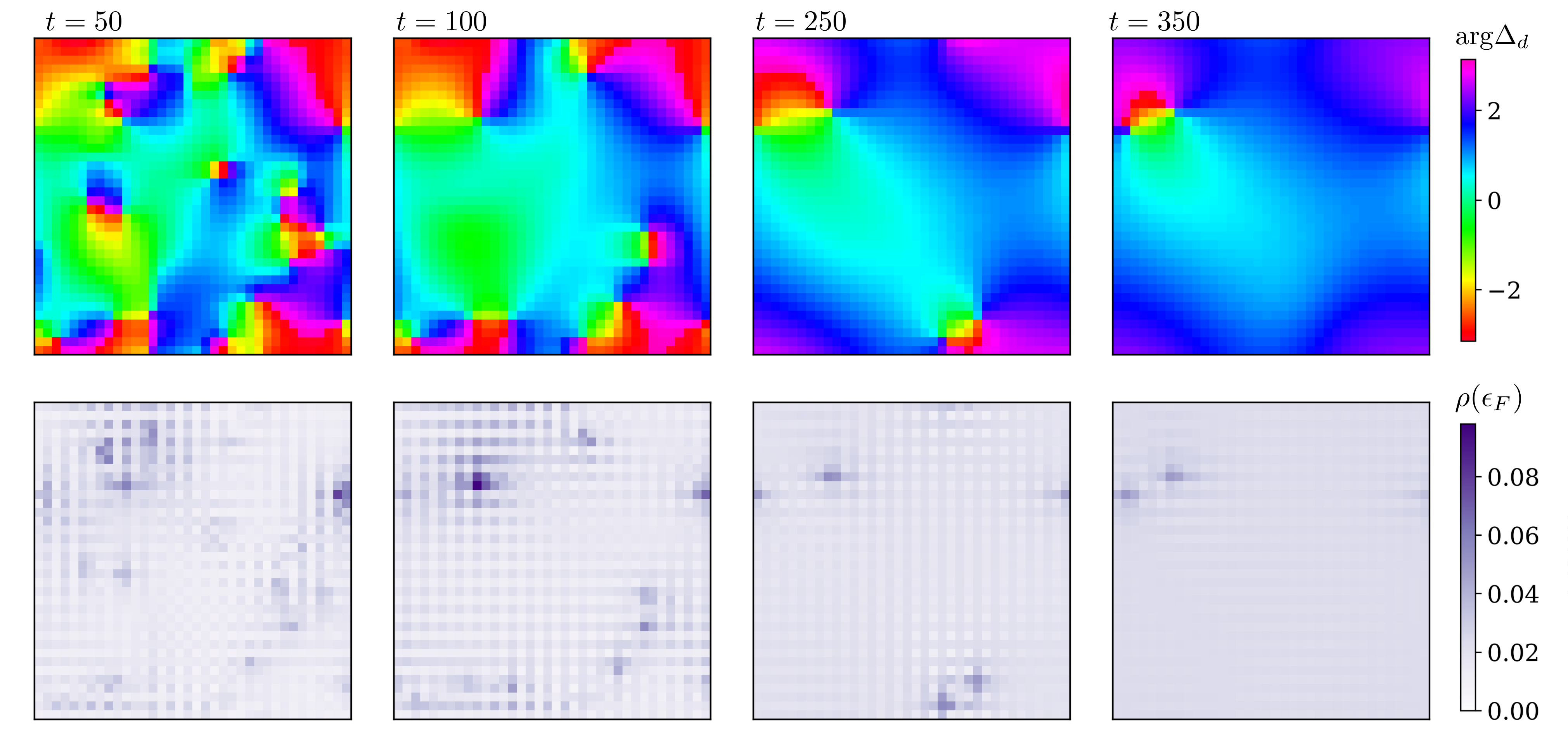}
\caption{Superconducting phase ordering and the accompanying electronic reconstruction. Columns show representative snapshots at $t=50$, $100$, $250$, and $350$, from left to right, with time measured in units of $t_0$. Top row: phase $\theta_i=\arg\Delta_i^d$ of the local $d$-wave pairing field. Vortices and antivortices appear as phase singularities and are identified by the integer plaquette winding described in Appendix~\ref{app:SC_vortices}. Bottom row: corresponding Fermi-level local density of states $\rho_i(\epsilon_F)$, with energy measured relative to the chemical potential. The phase texture coarsens as vortices and antivortices annihilate, while enhanced Fermi-level spectral weight is concentrated near the remaining vortex cores.}
\label{fig:SC_coarsening}
\end{figure*}

Numerically, we iterate the combined effective potentials $\Delta_{ij}^{\rm eff}$ directly, diagonalizing the BdG matrix and adjusting these potentials until the thermal pairing expectations reproduce the prescribed $\Delta_{ij}$. At convergence, $\Delta_{ij}^{\rm eff}=\Delta_{ij}-(V/2)f_{ij}$, so the thermodynamic forces are recovered from the difference between the prescribed and effective pairing fields. Stationarity of the constrained variational problem gives
\begin{equation}
f_{ij}=-\frac{\partial F}{\partial\Delta_{ij}^*}=\frac2V\left(\Delta_{ij}-\Delta_{ij}^{\rm eff}\right),
\label{eq:SC_force_recovery}
\end{equation}
with $\Delta_{ij}$ and $\Delta_{ij}^*$ treated independently. The Lagrange multipliers are therefore the negative free-energy derivatives and supply the thermodynamic forces without evaluating or numerically differentiating $F$. They vanish at an unconstrained equilibrium solution, where the ordinary BdG gap condition is recovered.

The real and imaginary pairing operators do not form a closed commutation algebra: their commutators generate density and hopping operators. We focus on purely relaxational dynamics, or model-A dynamics~\cite{Hohenberg1977}, of the nonconserved pairing fields, omitting the reversible contribution involving these additional expectation values. The deterministic equation of motion is
\begin{equation}
\frac{d\Delta_{ij}}{dt}=-\Gamma\frac{\partial F}{\partial\Delta_{ij}^*}=\Gamma f_{ij},
\label{eq:SC_modelA}
\end{equation}
where $\Gamma>0$ is a phenomenological kinetic coefficient with units of energy per time. We define $t_0=V/(2\Gamma)$ as the time unit. This evolution decreases the constrained free energy $F$ along a smooth branch of electronic states. Throughout the relaxation, the constrained BdG problem is solved at each instantaneous bond configuration, maintaining consistency between the evolving pairing fields and their microscopic expectation values~\footnote{See Appendices~\ref{app:SC_BdG_details}--\ref{app:SC_vortices} for the constrained mean-field functional, BdG response and LDOS, dynamical implementation and bond algebra, and vortex-identification procedure.}.

We now apply the constrained AVP dynamics to superconducting phase ordering on a $36\times36$ square lattice with periodic boundary conditions. We initialize the bond fields with spatially disordered complex values and follow their relaxation at fixed electronic temperature, using $V=4t$, $T=0.05t$, and $\mu=0$. The real and imaginary parts of the initial pairing fields are sampled independently from Gaussian distributions, without imposing a local $d$-wave relation between horizontal and vertical bonds. Energies are measured in units of the hopping $t$, and time in units of $t_0$.

Figure~\ref{fig:SC_coarsening} shows representative snapshots of the pairing phase $\theta_i=\arg\Delta_i^d$ and the corresponding Fermi-level LDOS. Local phase coherence develops rapidly: already at $t=50$, the earliest snapshot shown, the phase varies smoothly within extended regions. Global coherence has not yet been established, however, as the texture contains numerous vortices and antivortices, around which the phase winds by $+2\pi$ and $-2\pi$, respectively. Subsequent relaxation proceeds through vortex--antivortex annihilation, reducing the defect population and allowing phase coherence to extend over progressively larger distances.

The accompanying LDOS maps reveal the electronic reconstruction associated with this evolving phase texture. The local spectral weight $\rho_i(\epsilon_F)$ at the Fermi energy is evaluated from the instantaneous constrained BdG eigenstates. Enhanced spectral weight is concentrated near the phase singularities. As vortices and antivortices annihilate, the LDOS background becomes more uniform, while localized enhancements persist near the remaining cores. This spatial correspondence connects collective phase ordering to its microscopic electronic signatures, both obtained from the same evolving constrained states.

\begin{figure}[t]
\centering
\includegraphics[width=\columnwidth]{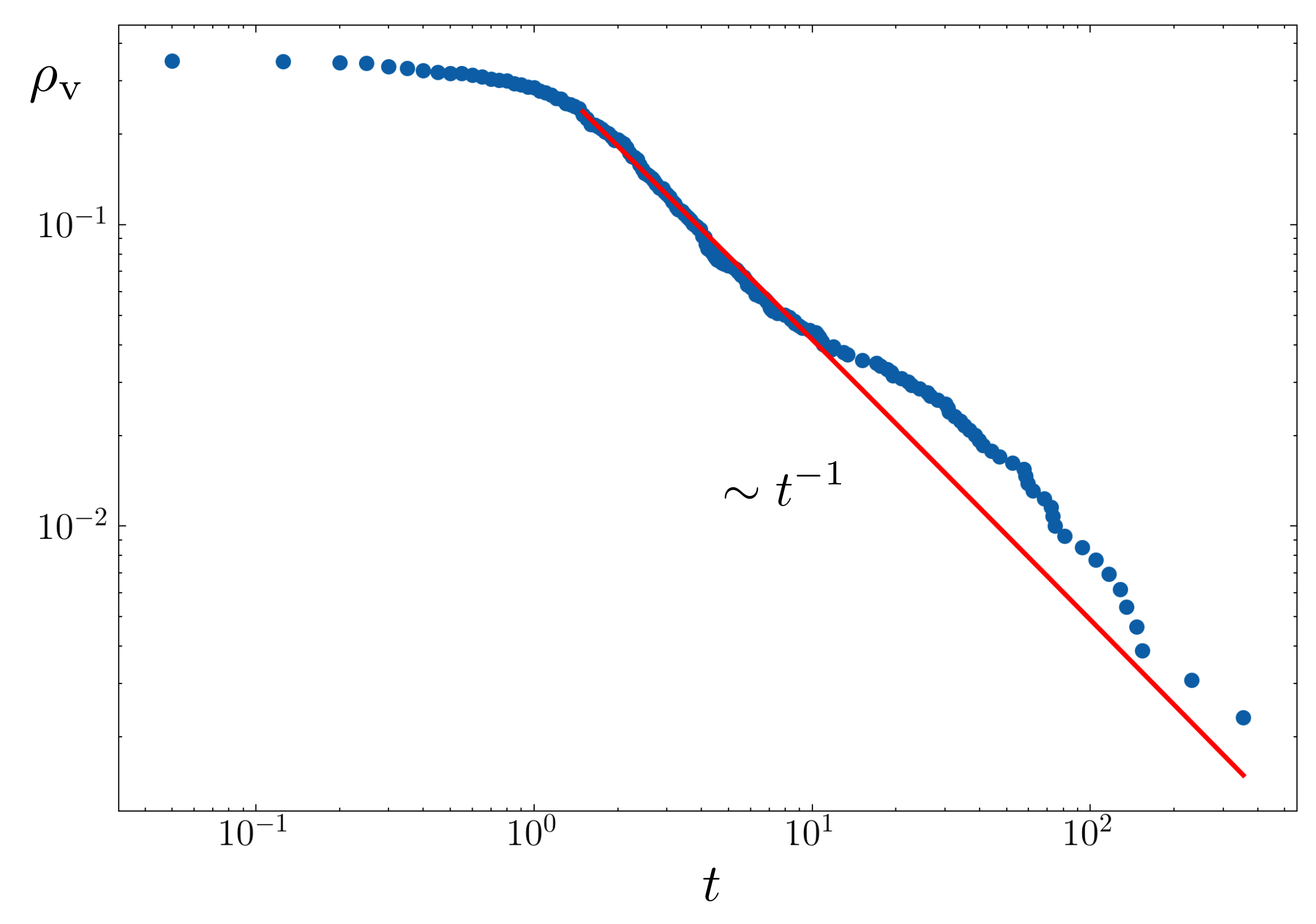}
\caption{Vortex density $\rho_v(t)$ during superconducting phase ordering, including both vortices and antivortices. Time is measured in units of $t_0$. Blue symbols show the simulation results; the red line indicates the reference behavior $t^{-1}$. An exploratory fit including a logarithmic correction is discussed in the text.}
\label{fig:SC_vortex_decay}
\end{figure}

We quantify the ordering kinetics through the vortex density $\rho_v(t)$, counting both vortices and antivortices. Figure~\ref{fig:SC_vortex_decay} shows an initial transient followed by an approximately inverse-time decay. To examine possible departures from this leading behavior, we consider the logarithmically corrected form
\begin{equation}
\rho_v(t)\simeq\frac{A}{t}\left[1+\gamma\ln t\right],
\label{eq:SC_vortex_decay_fit}
\end{equation}
where $A$ is an amplitude and $t$ is expressed in the time units specified above. An exploratory fit gives a small positive coefficient, $\gamma\simeq0.05$, although this estimate remains preliminary given the modest lattice size and limited scaling window. Such a correction is consistent with the established kinetics of two-dimensional phase vortices: the logarithmic vortex--antivortex attraction gives inverse-time density decay for an approximately constant mobility, while the additional logarithmic growth of vortex drag slows annihilation and produces the asymptotic behavior $\rho_v(t)\sim\ln(t)/t$ \cite{Yurke1993,Jelic2011}. Our results support approximately inverse-time decay with an indication of weak logarithmic slowing, rather than a precise determination of the asymptotic correction.

\begin{figure}[t]
\centering
\includegraphics[width=\columnwidth]{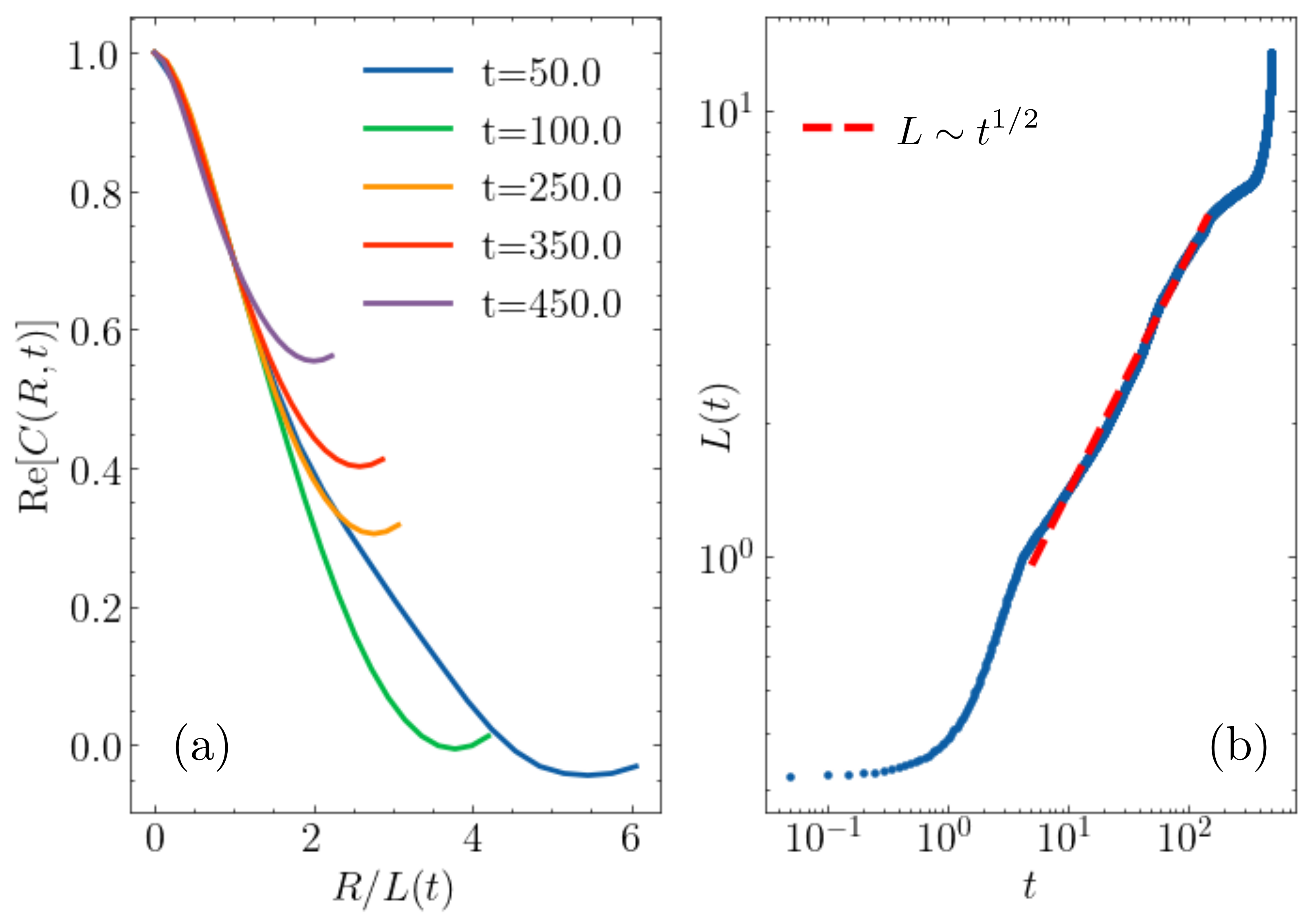}
\caption{Dynamical scaling of superconducting correlations. (a) Normalized pairing correlations at $t=50$, $100$, $250$, $350$, and $450$, plotted against $r/L(t)$. Approximate collapse occurs at smaller scaled separations, with larger-distance deviations becoming more pronounced at late times. (b) Extracted growing length $L(t)$. The red dashed line indicates the reference growth law $L(t)\sim t^{1/2}$. Times are measured in units of $t_0$.}
\label{fig:SC_dynamic_scaling}
\end{figure}

Spatial correlations provide a complementary characterization. Figure~\ref{fig:SC_dynamic_scaling}(a) shows the real part of the normalized equal-time pairing correlation $C(r,t)$ plotted against $r/L(t)$. The approximate collapse at smaller scaled separations supports the emergence of dynamical scaling, $\operatorname{Re}C(r,t)\simeq\mathcal C[r/L(t)]$. Deviations at larger separations, particularly at late times, are consistent with finite-size effects as the growing coherence length approaches the lattice dimensions. The extracted length in Fig.~\ref{fig:SC_dynamic_scaling}(b) is consistent with the leading diffusive law $L(t)\sim t^{1/2}$. In two dimensions, $\rho_v(t)\sim L(t)^{-2}$, connecting this square-root growth to the leading inverse-time vortex decay; the weak logarithmic correction corresponds to a small slowing of the length growth. Together, the defect decay and correlation scaling provide a nontrivial benchmark of the microscopic framework against established phase-ordering kinetics \cite{Bray1994,Yurke1993,Jelic2011}.

In summary, we have simulated dissipative superconducting phase ordering directly from the extended Hubbard Hamiltonian for the first time. The AVP construction enforces consistency between the evolving pairing fields and constrained electronic states, with Lagrange multipliers supplying the thermodynamic forces and BdG eigenstates resolving the accompanying local spectra. Vortex decay and correlation scaling consistent with conventional two-dimensional coarsening provide a nontrivial benchmark: collective vortex kinetics emerge from a microscopic thermodynamic landscape without prescribing a phenomenological Ginzburg--Landau functional.

Beyond sudden thermal quenches, AVP can address other bath-coupled protocols, including superconducting recovery in pump--probe experiments and cooling-rate-dependent defect formation in Kibble--Zurek annealing \cite{Kibble1976,Zurek1985,delCampo2014,Monaco2002,Maniv2003,Golubchik2010}. These applications require electronic equilibration to remain fast relative to collective evolution; following optical excitation, the adiabatic description would apply to recovery after rapid quasiparticle relaxation. With appropriate dissipative and fluctuation terms, the framework could address disorder-induced pairing inhomogeneity, pinning and dynamical arrest of vortices, and disorder-induced or self-generated pairing glasses, resolving microscopic barriers to superconducting coherence \cite{Ghosal2000,Fisher1991,Reichhardt2020,Mross2015,Yang2026a}.

The solver-independent formulation permits more correlated electronic states beyond the present Gaussian approximation. Its principal computational bottleneck, the repeated constrained electronic calculation, also provides a natural target for machine-learning acceleration. Building on advances in atomistic force fields and lattice-Hamiltonian simulations \cite{behler07,Drautz2019,Batzner2022,Batatia2022,zhang21,Cheng2023,Fan24,Fan26}, constrained AVP solutions could train models to predict thermodynamic forces directly from pairing configurations while preserving global phase covariance and lattice symmetries. Such models could reduce the need for iterative electronic solves and enable larger defect networks and longer relaxation times.

A particularly important direction is the coupled evolution of superconductivity and stripe or nematic order. Retaining pairing, charge, spin, and nematic fields together would permit microscopic studies of intertwined relaxation, including spatially modulated pair-density-wave states. Such simulations could determine how charge or nematic domains guide vortex motion, how competing orders develop in vortex halos, and whether coupled defects trap the system before global coherence is established \cite{Tranquada1995,Kivelson1998,Kivelson2003,Fradkin2010,Fernandes2014,Fernandes2019,Dagotto2005,Fradkin2015,Berg2007,Hamidian2016,Agterberg2020,Edkins2019}. This would connect the microscopic competition among electronic orders to the formation, recovery, and metastability of superconducting order.

\begin{acknowledgments}
This work was supported by the US Department of Energy Basic Energy Sciences under Contract No. DE-SC0020330. The authors acknowledge Research Computing at The University of Virginia for providing computational resources and technical support that have contributed to the results reported within this publication. 
\end{acknowledgments}

\onecolumngrid
\appendix

\section{Pairing channels and the half-filled superconducting state}
\label{app:SC_pairing_channels}

The nearest-neighbor density attraction does not select a spin-singlet channel by itself. With the unnormalized singlet operator $\hat B_{ij}$ defined in Eq.~\eqref{eq:SC_bond_order}, introduce the triplet vector $\hat{\boldsymbol T}_{ij}=(\sqrt{2}\hat c_{j\uparrow}\hat c_{i\uparrow},\hat c_{j\downarrow}\hat c_{i\uparrow}+\hat c_{j\uparrow}\hat c_{i\downarrow},\sqrt{2}\hat c_{j\downarrow}\hat c_{i\downarrow})^T$. For distinct sites, fermionic anticommutation gives the exact identity
\begin{equation}
\hat n_i\hat n_j=\frac12\left(\hat B_{ij}^\dagger\hat B_{ij}+\hat{\boldsymbol T}_{ij}^\dagger\!\cdot\hat{\boldsymbol T}_{ij}\right).
\label{eq:app_channel_decomposition}
\end{equation}
Thus the interaction contains $-(V/2)\hat B_{ij}^\dagger\hat B_{ij}$ and the corresponding triplet contribution with the same coupling. Retaining the singlet anomalous channel is a mean-field restriction, motivated here by its favorable pairing instability near half filling. Mean-field studies of the extended Hubbard model find $d_{x^2-y^2}$ pairing favored near half filling, with $p$-wave and extended-$s$ pairing becoming competitive as the filling is reduced \cite{Micnas1988,Micnas1990}. Related weak-coupling calculations also find competition between $d$-wave and chiral $p$-wave superconductivity as the Fermi surface changes with doping \cite{Huang2013}.

To make the channel comparison explicit for the normal-state band used in our implementation, take $\xi_{\boldsymbol k}=-2t(\cos k_x+\cos k_y)-\mu$ and the equally normalized form factors
\begin{equation}
g_d=\frac{\cos k_x-\cos k_y}{\sqrt{2}},\qquad g_s=\frac{\cos k_x+\cos k_y}{\sqrt{2}},\qquad g_{p_x}=\sin k_x,\qquad g_{p_y}=\sin k_y,
\label{eq:app_form_factors}
\end{equation}
whose Brillouin-zone averages satisfy $N^{-1}\sum_{\boldsymbol k}|g_\nu|^2=1/2$. Here $s$ denotes the extended-$s$ bond channel, rather than onsite pairing. Projecting the linearized pairing equation onto a channel gives
\begin{equation}
\chi_\nu(T)=\frac1N\sum_{\boldsymbol k}|g_\nu(\boldsymbol k)|^2\frac{\tanh[\xi_{\boldsymbol k}/(2T)]}{2\xi_{\boldsymbol k}},\qquad 2V\chi_\nu(T_c)=1,
\label{eq:app_susceptibility}
\end{equation}
where the ratio at $\xi_{\boldsymbol k}=0$ is understood as $1/(4T)$. The common factor $2V$ follows from the once-per-undirected-bond interaction and the normalization in Eq.~\eqref{eq:app_form_factors}; it is essential when comparing singlet and triplet channels.

At half filling of the fixed normal-state band, $\mu=0$. The $d$-wave form factor is nonzero at the van Hove points $(\pi,0)$ and $(0,\pi)$, whereas the $p$-wave form factors vanish there. Moreover, $g_s$ vanishes on the entire half-filled Fermi surface. Consequently, the low-temperature $d$-wave susceptibility benefits from the van Hove enhancement of the Cooper singularity, while the other channels receive less singular contributions. Numerical Brillouin-zone integration at $T/t=0.05$ gives $\chi_d/\chi_{p_x}=2.1614$, with the values in Table~\ref{tab:SC_channels}.

The susceptibility tests the instability of the normal state. A complementary comparison below the pairing transition minimizes the uniform pairing grand potential within each pure channel. For a quasiparticle gap $\Delta_{\boldsymbol k}=\delta g_\nu(\boldsymbol k)$, the grand-potential difference per site relative to the unpaired state is
\begin{equation}
\frac{F_\nu-F_{\rm normal}}{N}=\frac{|\delta|^2}{2V}-\frac{2T}{N}\sum_{\boldsymbol k}\ln\!\left[\frac{\cosh\!\left(\sqrt{\xi_{\boldsymbol k}^2+|\delta g_\nu(\boldsymbol k)|^2}/(2T)\right)}{\cosh[\xi_{\boldsymbol k}/(2T)]}\right].
\label{eq:app_uniform_free_energy}
\end{equation}
For the chiral triplet comparison, $g_{p_x+ip_y}=(\sin k_x+i\sin k_y)/\sqrt{2}$ has the same normalization. At $V/t=4$ and $T/t=0.05$, the $d$-wave state has the lowest grand potential among the uniform pure pairing states compared in Table~\ref{tab:SC_channels}. These calculations justify the superconducting sector studied here; they do not establish the unrestricted ground state of the full density-interaction Hamiltonian against charge ordering or phase separation. Normal Hartree and Fock renormalizations are omitted throughout the pairing-channel implementation, so $\mu=0$ refers to its fixed, half-filled band.

\begin{table}[htb]
\centering
\begin{tabular}{lcc}
\hline\hline
Pairing channel & $t\chi_\nu$ & $(F_\nu-F_{\rm normal})/(Nt)$ \\
\hline
Singlet $d_{x^2-y^2}$ & $0.852926$ & $-0.206139$ \\
Singlet extended $s$ & $0.101185$ & $0$ \\
Triplet $p_x$ or $p_y$ & $0.394615$ & $-0.145948$ \\
Triplet $p_x+ip_y$ & $0.394615$ & $-0.161517$ \\
\hline\hline
\end{tabular}
\caption{Pairing-channel comparison for $\mu=0$ and $T/t=0.05$. The second column is the normal-state susceptibility in Eq.~\eqref{eq:app_susceptibility}. The last column is the minimized grand-potential difference per site at $V/t=4$, in units of $t$. Brillouin-zone sums were evaluated on a $1200\times1200$ midpoint grid; the quoted digits are unchanged on an $800\times800$ grid. The extended-$s$ channel has no nonzero pure-channel solution at these parameters.}
\label{tab:SC_channels}
\end{table}

\section{Constrained mean-field theory and electronic observables}
\label{app:SC_BdG_details}

\subsection{Constrained free energy}
The evolving fields $\Delta_{ij}$ prescribe anomalous expectation values through Eq.~\eqref{eq:SC_bond_order}. The mean-field AVP minimizes the pairing-channel grand-potential functional over normalized Gaussian density operators $\hat\rho_{\rm G}$ subject to those constraints:
\begin{equation}
\begin{aligned}
F[\Delta]&=\min_{\substack{\hat\rho_{\rm G}\\\langle\hat B_{ij}\rangle=-(2/V)\Delta_{ij}}}F_{\rm MF}[\hat\rho_{\rm G}],\\
F_{\rm MF}[\hat\rho_{\rm G}]&=-t\sum_{\langle ij\rangle,\sigma}\langle\hat c_{i\sigma}^\dagger\hat c_{j\sigma}+{\rm h.c.}\rangle-\mu\sum_i\langle\hat n_i\rangle-\frac V2\sum_{\langle ij\rangle}|\langle\hat B_{ij}\rangle|^2+T\,{\rm Tr}(\hat\rho_{\rm G}\ln\hat\rho_{\rm G}).
\end{aligned}
\label{eq:app_constrained_F}
\end{equation}
All expectation values in the functional are evaluated in the trial state. In particular, they must not be replaced by their target values before varying that state. The Lagrangian in Eq.~\eqref{eq:SC_constrained_functional} imposes the real and imaginary parts of each constraint. Its stationary Gaussian state is the thermal state of Eq.~\eqref{eq:SC_BdG_Hamiltonian}, with $\Delta_{ij}^{\rm eff}=-(V/2)(\langle\hat B_{ij}\rangle+f_{ij})$. At convergence, the anomalous expectation equals its target and the relation simplifies to $\Delta_{ij}^{\rm eff}=\Delta_{ij}-(V/2)f_{ij}$.

The envelope identity is $dF=-\sum_{\langle ij\rangle}(f_{ij}^*d\Delta_{ij}+f_{ij}d\Delta_{ij}^*)$, yielding $f_{ij}=-\partial F/\partial\Delta_{ij}^*$ as in Eq.~\eqref{eq:SC_force_recovery}. Thus the Lagrange multipliers are precisely the thermodynamic forces driving Eq.~\eqref{eq:SC_modelA}. Since both $F$ and $\Delta_{ij}$ have energy units, $f_{ij}$ is dimensionless. For $\Delta_{ij}=x_{ij}+iy_{ij}$, we use $\partial/\partial\Delta_{ij}^*=(\partial/\partial x_{ij}+i\partial/\partial y_{ij})/2$, so $\partial F/\partial x_{ij}=-2\,{\rm Re}\,f_{ij}$ and $\partial F/\partial y_{ij}=-2\,{\rm Im}\,f_{ij}$. These identities fix the sign and factors entering the dynamics.

For evaluating $F$ numerically, let $F_{\rm BdG}[\Delta^{\rm eff}]=-T\ln{\rm Tr}\exp[-\hat{\mathcal H}_{\rm BdG}(\Delta^{\rm eff})/T]$ denote the grand potential of the auxiliary quadratic operator. In terms of its positive-energy eigenvalues, $F_{\rm BdG}={\rm Tr}\,h-2T\sum_{E_\nu>0}\ln[2\cosh(E_\nu/2T)]$; the ${\rm Tr}\,h$ term retains the normal-state constant in the reduced Nambu representation. At a converged constrained solution,
\begin{equation}
F[\Delta]=F_{\rm BdG}[\Delta^{\rm eff}]+\frac2V\sum_{\langle ij\rangle}\left[(\Delta_{ij}^{\rm eff})^*\Delta_{ij}+\Delta_{ij}^{\rm eff}\Delta_{ij}^*-|\Delta_{ij}|^2\right].
\label{eq:app_F_evaluation}
\end{equation}
The correction removes the auxiliary pairing-source energy and restores the physical mean-field interaction energy. At zero multiplier it reduces to the familiar condensation-energy correction $(2/V)\sum_{\langle ij\rangle}|\Delta_{ij}|^2$. A source-only quadratic grand potential, without this correction, is therefore insufficient to monitor the constrained free energy away from equilibrium. When several constrained stationary solutions exist, the force identities apply along the smooth branch being followed.

\subsection{BdG eigenstates and pairing response}
In the reduced Nambu basis $(\hat c_{1\uparrow},\ldots,\hat c_{N\uparrow},\hat c_{1\downarrow}^\dagger,\ldots,\hat c_{N\downarrow}^\dagger)^T$, the matrix of the constrained quadratic operator and its eigenproblem are
\begin{equation}
H_{\rm BdG}[\Delta^{\rm eff}]=\begin{pmatrix}h&\Delta^{\rm eff}\\(\Delta^{\rm eff})^\dagger&-h^T\end{pmatrix},\qquad h_{ij}=-tA_{ij}-\mu\delta_{ij},\qquad H_{\rm BdG}\begin{pmatrix}u_\nu\\v_\nu\end{pmatrix}=E_\nu\begin{pmatrix}u_\nu\\v_\nu\end{pmatrix}.
\label{eq:app_BdG_matrix}
\end{equation}
Here $A_{ij}$ is the nearest-neighbor adjacency matrix, and the symmetric matrix $\Delta^{\rm eff}$ contains the effective potentials on nearest-neighbor bonds. Eigenvectors have unit norm, $\sum_i[|u_\nu(i)|^2+|v_\nu(i)|^2]=1$. The spin-independent singlet problem has paired positive and negative energies. Thermal occupations are $n_F(E)=[\exp(E/T)+1]^{-1}$, with the chemical potential already included in $h$.

To distinguish the input pairing potential from its calculated anomalous response, define
\begin{equation}
D_{ij}[\Delta^{\rm eff}]\equiv-\frac V2\langle\hat B_{ij}\rangle_{\Delta^{\rm eff}}=\frac V2\sum_{E_\nu>0}\left[u_\nu(i)v_\nu^*(j)+u_\nu(j)v_\nu^*(i)\right]\tanh\!\left(\frac{E_\nu}{2T}\right).
\label{eq:app_pair_response}
\end{equation}
The subscript on the expectation value indicates the thermal state generated by the specified effective potential. No extra spin-degeneracy factor is inserted in this formula. The inner electronic calculation solves the inverse problem
\begin{equation}
D_{ij}[\Delta^{\rm eff}]=\Delta_{ij}.
\label{eq:app_inverse_constraint}
\end{equation}
The target on the right is the fixed instantaneous order, rather than the trial effective potential. Ordinary equilibrium self-consistency instead sets the response equal to the same field that enters the BdG Hamiltonian. The two conditions coincide only when $f_{ij}=0$. The inverse problem need only be solved on the constrained branch explored by the dynamics; global uniqueness or invertibility is not assumed.

\subsection{Local pairing symmetry}
The dynamical field is stored on bonds, and its site-centered $d$-wave projection is Eq.~\eqref{eq:SC_local_d}. The corresponding extended-$s$ component is
\begin{equation}
\Delta_i^s=\frac14\left(\Delta_{i,i+\hat x}+\Delta_{i,i-\hat x}+\Delta_{i,i+\hat y}+\Delta_{i,i-\hat y}\right).
\label{eq:app_local_s}
\end{equation}
We use superscripts $d$ and $s$ to label the local pairing symmetry. A uniform state with horizontal and vertical bonds $\Delta_0$ and $-\Delta_0$ has $\Delta_i^d=\Delta_0$ and $\Delta_i^s=0$, whereas equal bond values give the extended-$s$ state. Both projections are calculated from the physical evolving $\Delta_{ij}$, rather than from $\Delta_{ij}^{\rm eff}$. Independent bond evolution permits local extended-$s$ admixture, including near strongly inhomogeneous textures, even though the analysis focuses on $d$-wave coherence.

\subsection{Local density of states}
The same instantaneous constrained eigenstates determine the electronic spectrum \cite{deGennes1966,Gygi1991,Wang1995}. For one spin component, the site-resolved LDOS is
\begin{equation}
\rho_i(\omega)=\sum_{E_\nu>0}\left[|u_\nu(i)|^2\delta_\eta(\omega-E_\nu)+|v_\nu(i)|^2\delta_\eta(\omega+E_\nu)\right],\qquad \delta_\eta(x)=\frac{\eta}{\pi(x^2+\eta^2)}.
\label{eq:app_LDOS}
\end{equation}
The frequency is measured relative to $\mu$, and $\eta>0$ is the spectral broadening. The normalization is $\int d\omega\,\rho_i(\omega)=1$ per spin; the spin-summed LDOS is twice this expression. No Fermi occupation factor multiplies the LDOS, which is a spectral observable rather than an occupied-state density. Exactly zero-energy modes, if present, require their zero-mode spectral weight to be included separately instead of being discarded by the positive-energy filter. The Fermi-level maps follow by evaluating Eq.~\eqref{eq:app_LDOS} at $\omega=0$. Enhanced low-energy spectral weight is interpreted together with the pairing amplitude and phase winding, since nodal quasiparticles also contribute in a $d$-wave superconductor.

\section{Dissipative dynamics and numerical implementation}
\label{app:SC_dynamics_implementation}

\subsection{Reduced dynamics and bond algebra}
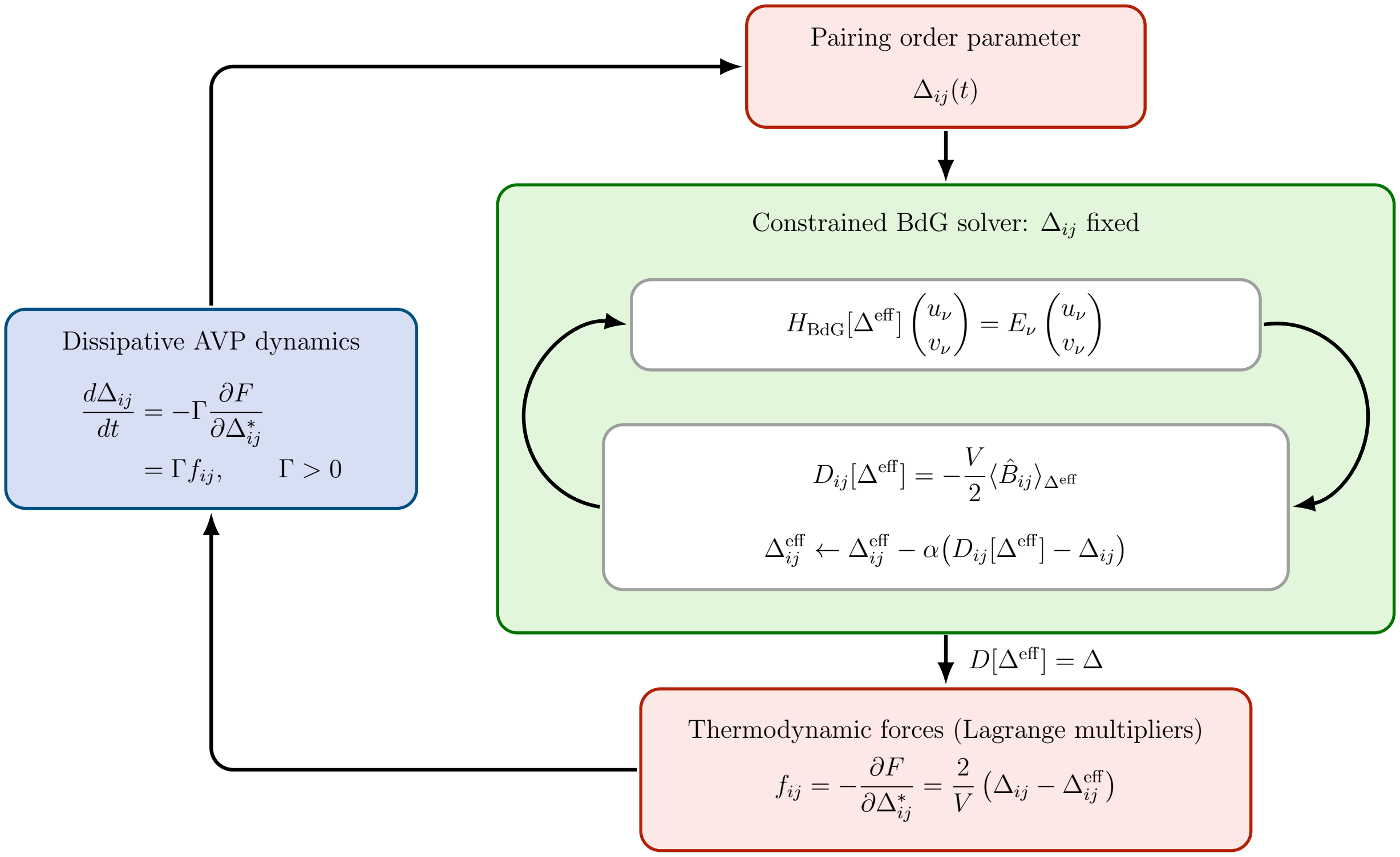
\begin{figure}[htb]
\centering
\resizebox{0.9\textwidth}{!}{%
\begin{tikzpicture}[x=1cm,y=1cm,>=Latex,font=\large,
 frame/.style={rounded corners=9pt,line width=1.4pt},
 box/.style={frame,draw=gray!75,fill=white,align=center},
 arrow/.style={->,line width=1.65pt,shorten >=1pt,shorten <=1pt}]

\node[box,draw=edgered,fill=outerpink,minimum width=6.4cm,minimum height=1.95cm] (order) at (7.2,14.2)
 {Pairing order parameter\\[10pt]$\displaystyle\Delta_{ij}(t)$};

\draw[frame,draw=edgegreen,fill=innergreen] (0,5.1) rectangle (14.4,12.3);
\node[align=center] at (7.2,11.65) {Constrained BdG solver: $\Delta_{ij}$ fixed};

\node[box,minimum width=10.1cm,minimum height=1.45cm] (eig) at (7.2,10.05)
 {$\displaystyle H_{\rm BdG}[\Delta^{\rm eff}]\begin{pmatrix}u_\nu\\v_\nu\end{pmatrix}
 =E_\nu\begin{pmatrix}u_\nu\\v_\nu\end{pmatrix}$};

\node[box,minimum width=11.0cm,minimum height=2.65cm] (response) at (7.2,7.12)
 {$\displaystyle D_{ij}[\Delta^{\rm eff}]=-\frac{V}{2}\langle\hat B_{ij}\rangle_{\Delta^{\rm eff}}$\\[14pt]
 $\displaystyle \Delta_{ij}^{\rm eff}\leftarrow\Delta_{ij}^{\rm eff}-\alpha\bigl(D_{ij}[\Delta^{\rm eff}]-\Delta_{ij}\bigr)$};

\draw[arrow]
 let \p1=(eig.east), \p2=(response.east),
     \n1={0.5*veclen(\x1-\x2,\y1-\y2)},
     \n2={atan2(\y1-\y2,\x1-\x2)}
 in (eig.east) arc[start angle=\n2,delta angle=-180,radius=\n1];

\draw[arrow]
 let \p1=(response.west), \p2=(eig.west),
     \n1={0.5*veclen(\x1-\x2,\y1-\y2)},
     \n2={atan2(\y1-\y2,\x1-\x2)}
 in (response.west) arc[start angle=\n2,delta angle=-180,radius=\n1];

\node[box,draw=edgered,fill=outerpink,minimum width=9.8cm,minimum height=2.6cm] (force) at (7.2,2.9)
 {Thermodynamic forces (Lagrange multipliers)\\[5pt]
 $\displaystyle f_{ij}=-\frac{\partial F}{\partial\Delta_{ij}^*}=\frac2V\left(\Delta_{ij}-\Delta_{ij}^{\rm eff}\right)$};

\node[box,draw=edgeblue,fill=outerblue,minimum width=6.6cm,minimum height=3.2cm] (dyn) at (-4.6,8.7)
 {Dissipative AVP dynamics\\[12pt]
 $\displaystyle\begin{aligned}\frac{d\Delta_{ij}}{dt}&=-\Gamma\frac{\partial F}{\partial\Delta_{ij}^*}\\&=\Gamma f_{ij},\qquad\Gamma>0\end{aligned}$};

\draw[arrow] (order.south) -- (7.2,12.3);
\draw[arrow] (7.2,5.1) -- (force.north);
\node[anchor=west,inner sep=3pt] at (7.45,4.65)
 {$D[\Delta^{\rm eff}]=\Delta$};

\draw[arrow,rounded corners=10pt] (force.west) -| (dyn.south);
\draw[arrow,rounded corners=10pt] (dyn.north) |- (order.west);

\end{tikzpicture}%
}
\caption{Constrained BdG implementation of superconducting AVP dynamics. The instantaneous pairing order parameter $\Delta_{ij}$ is supplied to the constrained BdG solver, whose inner iteration adjusts $\Delta_{ij}^{\rm eff}$ until the thermal pairing response reproduces the prescribed configuration. The converged Lagrange multipliers $f_{ij}=(2/V)(\Delta_{ij}-\Delta_{ij}^{\rm eff})=-\partial F/\partial\Delta_{ij}^*$ are the thermodynamic forces entering the dissipative equation of motion. The resulting evolution updates the pairing fields, completing the outer dynamical loop. The inner mixing parameter $\alpha$ belongs to the constraint solver; the physical time unit is $t_0=V/(2\Gamma)$.}
\label{fig:SC_AVP_flowchart}
\end{figure}

The electronic problem is solved at fixed temperature for each instantaneous bond configuration. Its eigenstates and thermal occupations are recomputed adiabatically, rather than propagated in real time. We retain only the deterministic dissipative dynamics of Eq.~\eqref{eq:SC_modelA}; the phenomenological coefficient $\Gamma$ defines the time unit $t_0=V/(2\Gamma)$. For a converged inner solve on a smooth constrained branch, Eq.~\eqref{eq:SC_force_recovery} implies
\begin{equation}
\frac{dF}{dt}=-2\Gamma\sum_{\langle ij\rangle}|f_{ij}|^2\leq0.
\label{eq:app_dissipation}
\end{equation}
The finite electronic temperature controls the constrained state. No stochastic force is included in the order-parameter evolution reported here. Figure~\ref{fig:SC_AVP_flowchart} summarizes the nested electronic and dynamical updates.

The restriction to dissipative evolution is distinct from the algebraic structure of the selected operators. Their commutators include
\begin{equation}
[\hat B_{ij},\hat B_{ij}^\dagger]=2-\hat n_i-\hat n_j,\qquad [\hat B_{ij},\hat B_{ik}^\dagger]=-\sum_\sigma\hat c_{k\sigma}^\dagger\hat c_{j\sigma}\quad(i,j,k\ {\rm distinct}).
\label{eq:app_bond_algebra}
\end{equation}
Consequently, the real and imaginary pairing quadratures generate density and hopping bilinears and do not constitute a closed algebra. In the general AVP formulation, a reversible contribution can involve the expectation values of these additional operators \cite{Yang2026b}. Nonclosure does not make that contribution vanish; omitting it here defines the purely relaxational dynamical model.

\subsection{Inverse iteration and outer time step}
At a fixed outer configuration, define the bond residual $R_{ij}^{(k)}=D_{ij}[\Delta^{{\rm eff},(k)}]-\Delta_{ij}$. The effective potential is adjusted by
\begin{equation}
\Delta_{ij}^{{\rm eff},(k+1)}=\Delta_{ij}^{{\rm eff},(k)}-\alpha_k R_{ij}^{(k)}.
\label{eq:app_inner_iteration}
\end{equation}
The implementation uses a Barzilai--Borwein-type step \cite{Barzilai1988}. Writing $s^{(k)}=\Delta^{{\rm eff},(k)}-\Delta^{{\rm eff},(k-1)}$ and $y^{(k)}=R^{(k)}-R^{(k-1)}$ as arrays over the independent bonds, the trial step is
\begin{equation}
\alpha_k^{\rm BB}=\frac{{\rm Re}\langle y^{(k)},s^{(k)}\rangle}{{\rm Re}\langle y^{(k)},y^{(k)}\rangle},\qquad \langle X,Y\rangle=\sum_{\langle ij\rangle}X_{ij}^*Y_{ij}.
\label{eq:app_BB_step}
\end{equation}
The step is clipped to $[-1,1]$, with initial value $1$; the previous step is reused if the denominator falls below $10^{-24}$ in the numerical units. This is a nonlinear root iteration, not a guaranteed descent of $F$ or of the residual norm. Convergence requires the maximum absolute real and imaginary residual components to be separately below $10^{-6}$; the maximum inner count is $50\,000$.

The response must be evaluated at the final accepted effective potential before extracting $f_{ij}$. Only a successful constrained solve supplies the force for an outer step. With rescaled time $\tau=t/t_0=2\Gamma t/V$, the explicit Euler update is
\begin{equation}
\Delta_{ij}(\tau+\delta\tau)=\Delta_{ij}(\tau)+\frac V2\delta\tau\,f_{ij}(\tau)=\Delta_{ij}(\tau)-\delta\tau\,[\Delta_{ij}^{\rm eff}(\tau)-\Delta_{ij}(\tau)].
\label{eq:app_outer_step}
\end{equation}
The dimensionless $\tau$ is denoted by $t$ in the main text and plots. The response-based update in the numerical implementation, with the bracket replaced by $\Delta_{ij}^{\rm eff}-D_{ij}[\Delta^{\rm eff}]$, agrees within the inner residual tolerance. The converged potential from one outer step provides an initial guess for the next. Euler integration is first order in $\delta\tau$; the continuous-time identity in Eq.~\eqref{eq:app_dissipation} does not by itself guarantee monotonicity for an arbitrary finite step. Inner residuals, time-step dependence, and the constrained free energy provide complementary numerical convergence checks.

\begin{table}[htb]
\centering
\begin{tabular}{ll}
\hline\hline
Parameter & Value \\
\hline
Lattice and boundaries & $36\times36$ ($N=1296$), periodic \\
Hopping and attraction & $t=1$, $V=4$ \\
Electronic temperature and chemical potential & $T=0.05$, $\mu=0$ \\
Outer time step & $\delta\tau=0.02$ \\
Number of outer steps; final rescaled time & $10\,000$; $\tau_{\rm final}=200$ \\
Inner residual tolerance & $10^{-6}$ per real/imaginary component \\
Maximum inner iterations & $50\,000$ \\
Initial inner step; clipping interval & $1$; $[-1,1]$ \\
Small-denominator threshold for BB step & $10^{-24}$ \\
GPU correlation-kernel threads per block & $256$ \\
\hline\hline
\end{tabular}
\caption{Baseline parameters of the documented Julia/CUDA implementation. Energies are measured in units of the hopping $t$, and the outer evolution uses $\tau=t/t_0=2\Gamma t/V$, with $t_0=V/(2\Gamma)$.}
\label{tab:SC_implementation}
\end{table}

\subsection{Storage, GPU evaluation, and baseline parameters}
For a periodic $N_x\times N_y$ lattice with $N=N_xN_y$, each undirected bond is stored once as $\Delta^x_i=\Delta_{i,i+\hat x}$ and $\Delta^y_i=\Delta_{i,i+\hat y}$. Backward bonds follow by index shifts, for example $\Delta_{i,i-\hat x}=\Delta^x_{i-\hat x}$. The effective potentials use the same storage. The Julia implementation \cite{Bezanson2017} assembles the complex Hermitian $2N\times2N$ BdG matrix, diagonalizes it on a GPU through CUDA and cuSOLVER \cite{Besard2019}, and uses a custom kernel to evaluate Eq.~\eqref{eq:app_pair_response} on the independent bonds. The kernel uses $256$ threads per block. Dense diagonalization costs $O(N^3)$ operations and $O(N^2)$ storage; evaluating the bond response from the eigenvectors costs $O(N^2)$ operations.

Table~\ref{tab:SC_implementation} lists the baseline implementation parameters documented in the numerical note. For $N=1296$, the reduced BdG matrix has dimension $2592$; one complex-double matrix occupies approximately $107.5$ MB in decimal units, excluding eigenvector copies and solver workspace. The initial bond targets are independent complex Gaussian variables, $\Delta_i^a(0)=[X_i^a+iY_i^a]/2$, where $a=x,y$ and the real variables are independent standard Gaussians. The initial trial effective potential is zero. As with later configurations, an initial target is usable only if its constrained electronic problem can be solved; arbitrary complex targets are not guaranteed to be representable by a Gaussian state. Complex site-centered $d$-wave fields are stored over time as arrays indexed by time and the two spatial coordinates, in NPZ-compatible output.

\begin{figure}[t]
\centering
\includegraphics[width=0.65\textwidth]{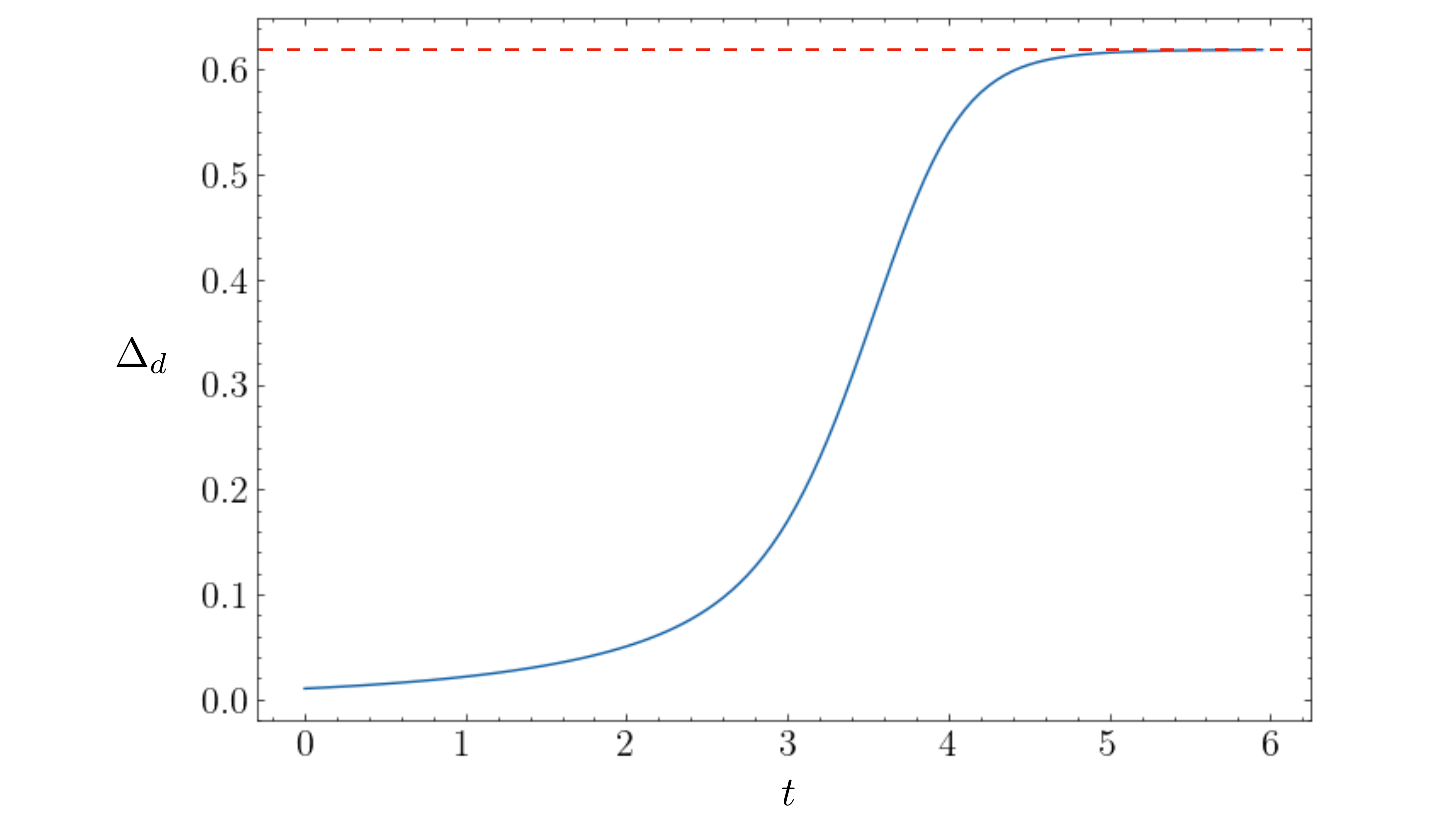}
\caption{Uniform-order benchmark of superconducting AVP dynamics. The bond fields are restricted to a spatially uniform $d$-wave configuration, with real $\Delta_i^d=\Delta^d(t)$ and opposite signs on horizontal and vertical bonds. The blue curve shows the growth and saturation of the pairing amplitude under purely relaxational AVP evolution; the red dashed line marks the standard self-consistent BdG ground-state value. Time is measured in units of $t_0=V/(2\Gamma)$ and the pairing amplitude in units of the hopping $t$.}
\label{fig:SC_uniform_benchmark}
\end{figure}

\begin{figure}[b]
\centering
\includegraphics[width=\textwidth]{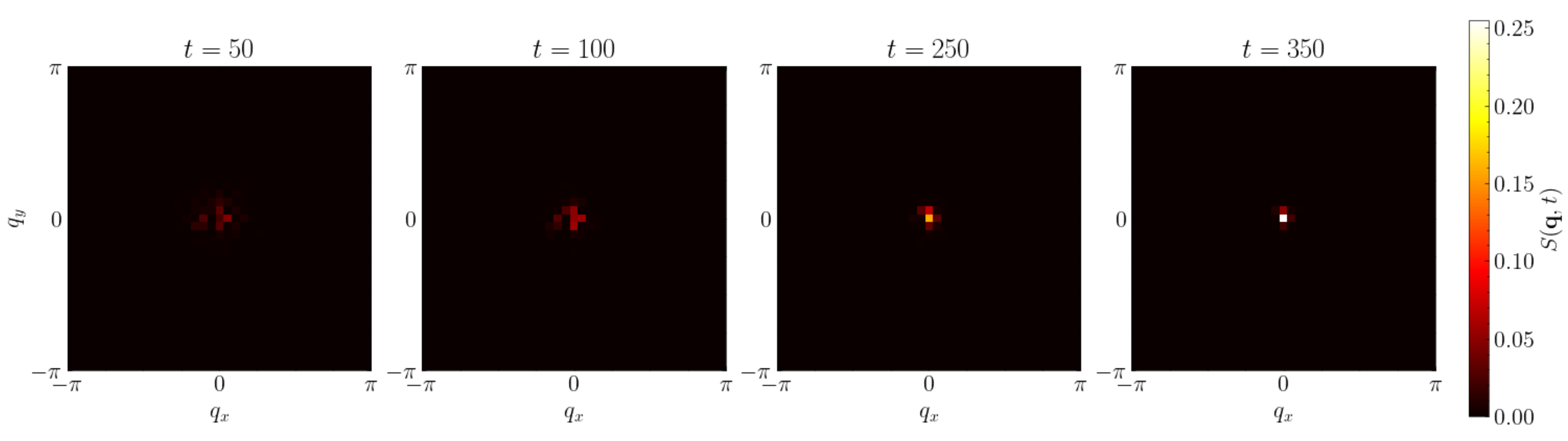}
\caption{Growth of superconducting coherence in momentum space. The panels show the structure factor $S(\mathbf q,t)$ of the local $d$-wave pairing field $\Delta_i^d$ at $t=50$, $100$, $250$, and $350$, from left to right, corresponding to the real-space snapshots in Fig.~\ref{fig:SC_coarsening}. The increasing height and decreasing width of the peak at $\mathbf q=0$ reflect the growth of phase-coherent regions toward global coherence. Momenta are shown in units of the inverse lattice spacing, and time is measured in units of $t_0$.}
\label{fig:SC_structure_factor}
\end{figure}

\subsection{Uniform-order benchmark and momentum-space coarsening}
With the numerical implementation specified, we present two additional results that complement the coarsening analysis in the main text. We first benchmark the constrained forces and their dynamical implementation by restricting the pairing field to a spatially uniform $d$-wave form, $\Delta_{i,i+\hat x}(t)=\Delta^d(t)$ and $\Delta_{i,i+\hat y}(t)=-\Delta^d(t)$, with $\Delta^d(t)$ chosen real. Starting from a small nonzero amplitude, the AVP evolution increases $\Delta^d$ and approaches the conventional self-consistent BdG ground-state value, shown by the red dashed line in Fig.~\ref{fig:SC_uniform_benchmark}. At the stationary point, the thermodynamic forces vanish and the constrained effective pairing potential coincides with the physical order parameter, recovering the ordinary gap equation. This uniform test verifies that the constrained dynamics recovers the expected microscopic stationary state.

Turning to the unrestricted, spatially inhomogeneous coarsening simulations discussed in the main text, the structure factor $S(\mathbf q,t)$ provides a complementary view of the development of phase coherence. It is obtained from the squared modulus of the spatial Fourier transform of the complex field $\Delta_i^d(t)$. Figure~\ref{fig:SC_structure_factor} shows its evolution at the same times as the real-space snapshots in Fig.~\ref{fig:SC_coarsening}. Although extended regions already have locally coherent phases at the earliest time shown, their relative phases remain disordered and the low-momentum weight is spread over several wave vectors. As vortices and antivortices annihilate, this weight becomes increasingly concentrated near $\mathbf q=0$: the peak grows and narrows as phase coherence extends over larger distances. The momentum-space evolution thus complements the real-space coarsening and correlation-length growth discussed in the main text.

\section{Phase symmetry and vortex identification}
\label{app:SC_vortices}

\begin{figure}[b]
\centering
\includegraphics[width=\textwidth]{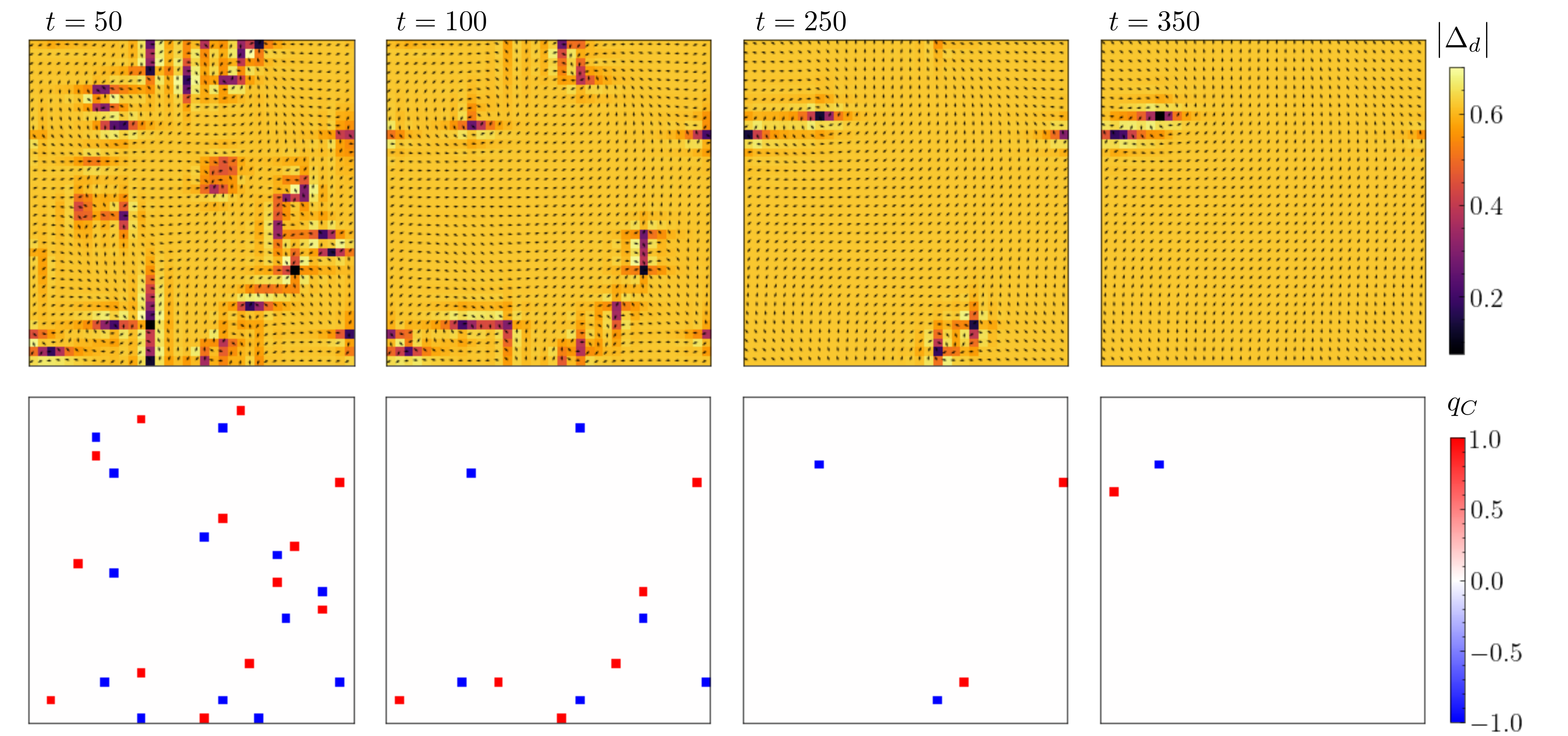}
\caption{Identification of vortices during superconducting phase ordering. Columns correspond to $t=50$, $100$, $250$, and $350$, from left to right, with time measured in units of $t_0$. Top row: the local pairing amplitude $|\Delta_i^d|$, with arrows indicating the phase of $\Delta_i^d$. Bottom row: the integer winding $q_C$ from Eq.~\eqref{eq:app_vortex_winding}, evaluated on counterclockwise elementary plaquettes. Red and blue mark vortices ($q_C=+1$) and antivortices ($q_C=-1$), respectively; white denotes zero winding. The identified defects coincide with regions of strongly suppressed pairing amplitude, and their number decreases as vortex--antivortex pairs annihilate.}
\label{fig:SC_vortex_identification}
\end{figure}

For a fixed $d_{x^2-y^2}$ spatial form, the remaining order-parameter freedom is its complex phase. Under a global fermionic transformation $\hat c_{i\sigma}\mapsto e^{i\varphi}\hat c_{i\sigma}$, both $\Delta_{ij}$ and $\Delta_i^d$ transform as $e^{2i\varphi}$. Away from a vortex core, write $\Delta_i^d(t)=|\Delta_i^d(t)|e^{i\theta_i(t)}$. The phase of the pair field takes values on $U(1)\simeq S^1$, and the relevant fundamental group is $\pi_1(S^1)=\mathbb Z$ \cite{Mermin1979}. A loop encircling a defect can therefore carry an integer phase winding. The fixed sign change between horizontal and vertical bonds specifies the $d$-wave form factor; it does not alter this integer classification of ordinary phase vortices.

For an oriented closed contour $C$ in the site-centered field, the winding number is
\begin{equation}
q_C=\frac1{2\pi}\sum_{(i\to j)\in C}{\rm Arg}\!\left[\Delta_j^d(\Delta_i^d)^*\right],\qquad {\rm Arg}\in(-\pi,\pi].
\label{eq:app_vortex_winding}
\end{equation}
This is the usual discrete phase-circulation prescription, with each phase difference reduced to its principal interval before summing \cite{Jelic2011,Bietenholz2013}. For a counterclockwise elementary plaquette, $q_p=+1$ identifies a vortex and $q_p=-1$ an antivortex; a regular plaquette has zero winding. Numerically, the sum divided by $2\pi$ is checked against the nearest integer. Simply summing unwrapped endpoint phases would telescope and miss the defect, while testing only for small $|\Delta_i^d|$ would confuse vortices with other amplitude suppressions.

Figure~\ref{fig:SC_vortex_identification} illustrates this identification for the same sequence of configurations shown in Fig.~\ref{fig:SC_coarsening}. The upper panels display $|\Delta_i^d|$ together with arrows indicating the local phase, while the lower panels show the corresponding plaquette windings. The nonzero charges are quantized to $+1$ and $-1$, identifying vortices and antivortices near local minima of the pairing amplitude. The amplitude is strongly suppressed in the core regions, where the phase winds rapidly, and recovers away from them. Because a core can lie between lattice sites, its presence need not produce an exactly vanishing value at a sampled site. The winding therefore supplies the defect identification, while the amplitude and phase textures provide a consistent spatial check. Their evolution shows the decreasing defect population underlying the phase ordering in the main text.

The plaquette center provides the lattice-scale defect position. If an amplitude vanishes on the chosen contour, its phase is undefined and the contour must be displaced or enlarged to pass through nonzero-amplitude sites. For poorly resolved cores, winding on larger enclosing contours provides a useful check. The elementary-plaquette construction assumes that the phase texture is resolved on the lattice scale; it cannot resolve an arbitrarily close vortex--antivortex pair lying inside one plaquette. With periodic boundary conditions and no imposed magnetic flux, the signed plaquette charges sum to zero for a nonsingular, resolved configuration, although noncontractible cycles of the torus may carry a global phase twist.

The total number of elementary defects and their density per plaquette are $N_v=\sum_p|q_p|$ and $n_v=N_v/(N_xN_y)$, respectively. Counting both signs measures the population that decays through vortex--antivortex annihilation during ordering \cite{Yurke1993,Jelic2011}; no additional separation into bound and free vortices is implicit in this count. The topological interpretation is the same one underlying vortex physics in two-dimensional $U(1)$ systems \cite{Kosterlitz1973}, without requiring an equilibrium BKT analysis of the present deterministic relaxation. Since the Hamiltonian used here has no electromagnetic gauge field or Peierls links, Eq.~\eqref{eq:app_vortex_winding} is evaluated directly from the pair field in the fixed lattice gauge. Including a spatial gauge field would require the corresponding gauge-covariant phase circulation and magnetic-flux contribution.

\clearpage
\twocolumngrid
\bibliography{ref}

@article{Mori1965,
  author = {H. Mori},
  title = {Transport, Collective Motion, and Brownian Motion},
  journal = {Progress of Theoretical Physics},
  volume = {33},
  pages = {423--455},
  year = {1965},
  doi = {10.1143/PTP.33.423}
}

@article{Zwanzig1960,
  author = {R. Zwanzig},
  title = {Ensemble Method in the Theory of Irreversibility},
  journal = {Journal of Chemical Physics},
  volume = {33},
  pages = {1338--1341},
  year = {1960},
  doi = {10.1063/1.1731409}
}

@article{Grabert1978,
  author    = {Hermann Grabert},
  title     = {Nonlinear Transport and Dynamics of Fluctuations},
  journal   = {Journal of Statistical Physics},
  volume    = {19},
  number    = {5},
  pages     = {479--497},
  year      = {1978},
  doi       = {10.1007/BF01011694}
}

@article{Hohenberg1977,
  title = {Theory of dynamic critical phenomena},
  author = {Hohenberg, P. C. and Halperin, B. I.},
  journal = {Rev. Mod. Phys.},
  volume = {49},
  issue = {3},
  pages = {435--479},
  numpages = {0},
  year = {1977},
  month = {Jul},
  publisher = {American Physical Society},
  doi = {10.1103/RevModPhys.49.435},
  url = {https://link.aps.org/doi/10.1103/RevModPhys.49.435}
}

@article{Bray1994,
  author = {A.J. Bray},
  title = {Theory of phase-ordering kinetics},
  journal = {Advances in Physics},
  volume = {43},
  number = {3},
  pages = {357--459},
  year = {1994},
  publisher = {Taylor \& Francis},
  doi = {10.1080/00018739400101505},
  url = {https://doi.org/10.1080/00018739400101505}
}

@article{Mermin1979,
  author    = {Mermin, N. D.},
  title     = {The topological theory of defects in ordered media},
  journal   = {Rev. Mod. Phys.},
  volume    = {51},
  number    = {3},
  pages     = {591--648},
  year      = {1979},
  month     = jul,
  publisher = {American Physical Society},
  doi       = {10.1103/RevModPhys.51.591},
  url       = {https://doi.org/10.1103/RevModPhys.51.591}
}

@article{Kibble1976,
  author = {Kibble, T. W. B.},
  title = {Topology of cosmic domains and strings},
  journal = {Journal of Physics A: Mathematical and General},
  volume = {9},
  pages = {1387--1398},
  year = {1976},
  doi = {10.1088/0305-4470/9/8/029},
  url = {https://doi.org/10.1088/0305-4470/9/8/029}
}

@article{Zurek1985,
  author = {Zurek, W. H.},
  title = {Cosmological experiments in superfluid helium?},
  journal = {Nature},
  volume = {317},
  pages = {505--508},
  year = {1985},
  doi = {10.1038/317505a0},
  url = {https://doi.org/10.1038/317505a0}
}

@article{delCampo2014,
  author = {del Campo, Adolfo and Zurek, Wojciech H.},
  title = {Universality of phase transition dynamics: Topological defects from symmetry breaking},
  journal = {International Journal of Modern Physics A},
  volume = {29},
  pages = {1430018},
  year = {2014},
  doi = {10.1142/s0217751x1430018x},
  url = {https://doi.org/10.1142/s0217751x1430018x}
}

@article{Dagotto2005,
  author  = {Dagotto, Elbio},
  title   = {Complexity in strongly correlated electronic systems},
  journal = {Science},
  volume  = {309},
  number  = {5732},
  pages   = {257--262},
  year    = {2005},
  doi     = {10.1126/science.1107559}
}

@article{Keimer2015,
  author  = {Keimer, B. and Kivelson, S. A. and Norman, M. R. and Uchida, S. and Zaanen, J.},
  title   = {From quantum matter to high-temperature superconductivity in copper oxides},
  journal = {Nature},
  volume  = {518},
  pages   = {179--186},
  year    = {2015},
  doi     = {10.1038/nature14165}
}

@article{Dagotto1994,
  author  = {Dagotto, Elbio},
  title   = {Correlated electrons in high-temperature superconductors},
  journal = {Rev. Mod. Phys.},
  volume  = {66},
  pages   = {763--840},
  year    = {1994},
  doi     = {10.1103/RevModPhys.66.763}
}

@article{Fan26,
    author = {Fan, Yunhao and Chern, Gia-Wei},
    title = {Graph neural network force fields for adiabatic dynamics of lattice Hamiltonians},
    journal = {APL Machine Learning},
    volume = {4},
    number = {3},
    pages = {036107},
    year = {2026},
    month = {07},
    issn = {2770-9019},
    doi = {10.1063/5.0334142},
    url = {https://doi.org/10.1063/5.0334142},
}

@article{Fan24,
  title = {Coarsening of chiral domains in itinerant electron magnets: A machine learning force-field approach},
  author = {Fan, Yunhao and Zhang, Sheng and Chern, Gia-Wei},
  journal = {Phys. Rev. B},
  volume = {110},
  issue = {24},
  pages = {245105},
  numpages = {18},
  year = {2024},
  month = {Dec},
  publisher = {American Physical Society},
  doi = {10.1103/PhysRevB.110.245105},
  url = {https://link.aps.org/doi/10.1103/PhysRevB.110.245105}
}

@article{Cheng2023,
    title = {Machine learning for phase ordering dynamics of charge density waves},
    author = {Cheng, Chen and Zhang, Sheng and Chern, Gia-Wei},
    journal = {Phys. Rev. B},
    volume = {108},
    issue = {1},
    pages = {014301},
    numpages = {15},
    year = {2023},
    month = {Jul},
    publisher = {American Physical Society},
    doi = {10.1103/PhysRevB.108.014301},
    url = {https://link.aps.org/doi/10.1103/PhysRevB.108.014301}
}

@article{zhang21,
  title = {Arrested Phase Separation in Double-Exchange Models: Large-Scale Simulation Enabled by Machine Learning},
  author = {Zhang, Puhan and Chern, Gia-Wei},
  journal = {Phys. Rev. Lett.},
  volume = {127},
  issue = {14},
  pages = {146401},
  numpages = {7},
  year = {2021},
  month = {Sep},
  publisher = {American Physical Society},
  doi = {10.1103/PhysRevLett.127.146401},
  url = {https://link.aps.org/doi/10.1103/PhysRevLett.127.146401}
}

@article{behler07,
  title = {Generalized Neural-Network Representation of High-Dimensional Potential-Energy Surfaces},
  author = {Behler, J\"org and Parrinello, Michele},
  journal = {Phys. Rev. Lett.},
  volume = {98},
  issue = {14},
  pages = {146401},
  numpages = {4},
  year = {2007},
  month = {Apr},
  publisher = {American Physical Society},
  doi = {10.1103/PhysRevLett.98.146401},
  url = {https://link.aps.org/doi/10.1103/PhysRevLett.98.146401}
}

@article{Batzner2022,
  author = {Batzner, Simon and Musaelian, Albert and Sun, Lixin and Geiger, Mario and Mailoa, Jonathan P. and Kornbluth, Mordechai and Molinari, Nicola and Smidt, Tess E. and Kozinsky, Boris},
  title = {{E(3)}-equivariant graph neural networks for data-efficient and accurate interatomic potentials},
  journal = {Nature Communications},
  volume = {13},
  pages = {2453},
  year = {2022},
  doi = {10.1038/s41467-022-29939-5},
  url = {https://doi.org/10.1038/s41467-022-29939-5}
}

@article{Fradkin2015,
  author    = {Fradkin, Eduardo and Kivelson, Steven A. and Tranquada, John M.},
  title     = {Colloquium: Theory of intertwined orders in high temperature superconductors},
  journal   = {Rev. Mod. Phys.},
  volume    = {87},
  number    = {2},
  pages     = {457--482},
  year      = {2015},
  doi       = {10.1103/RevModPhys.87.457},
  url       = {https://doi.org/10.1103/RevModPhys.87.457},
}

@article{Kivelson2003,
  author    = {Kivelson, S. A. and Bindloss, I. P. and Fradkin, E. and Oganesyan, V. and Tranquada, J. M. and Kapitulnik, A. and Howald, C.},
  title     = {How to detect fluctuating stripes in the high-temperature superconductors},
  journal   = {Rev. Mod. Phys.},
  volume    = {75},
  number    = {4},
  pages     = {1201--1241},
  year      = {2003},
  doi       = {10.1103/RevModPhys.75.1201},
  url       = {https://doi.org/10.1103/RevModPhys.75.1201},
}

@article{Kivelson1998,
  author    = {Kivelson, S. A. and Fradkin, E. and Emery, V. J.},
  title     = {Electronic liquid-crystal phases of a doped {Mott} insulator},
  journal   = {Nature},
  volume    = {393},
  number    = {6685},
  pages     = {550--553},
  year      = {1998},
  doi       = {10.1038/31177},
  url       = {https://doi.org/10.1038/31177},
}

@article{Fradkin2010,
  author    = {Fradkin, Eduardo and Kivelson, Steven A. and Lawler, Michael J. and Eisenstein, James P. and Mackenzie, Andrew P.},
  title     = {Nematic {Fermi} Fluids in Condensed Matter Physics},
  journal   = {Annu. Rev. Condens. Matter Phys.},
  volume    = {1},
  number    = {1},
  pages     = {153--178},
  year      = {2010},
  doi       = {10.1146/annurev-conmatphys-070909-103925},
  url       = {https://doi.org/10.1146/annurev-conmatphys-070909-103925},
}

@article{Fernandes2019,
  author    = {Fernandes, Rafael M. and Orth, Peter P. and Schmalian, J{\"o}rg},
  title     = {Intertwined Vestigial Order in Quantum Materials: Nematicity and Beyond},
  journal   = {Annu. Rev. Condens. Matter Phys.},
  volume    = {10},
  number    = {1},
  pages     = {133--154},
  year      = {2019},
  doi       = {10.1146/annurev-conmatphys-031218-013200},
  url       = {https://doi.org/10.1146/annurev-conmatphys-031218-013200},
}

@article{Agterberg2020,
  author    = {Agterberg, Daniel F. and Davis, J.C. S{\'e}amus and Edkins, Stephen D. and Fradkin, Eduardo and Van Harlingen, Dale J. and Kivelson, Steven A. and Lee, Patrick A. and Radzihovsky, Leo and Tranquada, John M. and Wang, Yuxuan},
  title     = {The Physics of Pair-Density Waves: Cuprate Superconductors and Beyond},
  journal   = {Annu. Rev. Condens. Matter Phys.},
  volume    = {11},
  number    = {1},
  pages     = {231--270},
  year      = {2020},
  doi       = {10.1146/annurev-conmatphys-031119-050711},
  url       = {https://doi.org/10.1146/annurev-conmatphys-031119-050711},
}

@article{Yang2026a,
  author        = {Yang, Lingyu and Chern, Gia-Wei},
  title         = {Electronic glasses from a broken gauge symmetry in disorder-free systems},
  journal       = {Physical Review B},
  volume        = {113},
  pages         = {L100201},
  year          = {2026},
  doi           = {10.1103/17sg-tx9k},
  url           = {https://doi.org/10.1103/17sg-tx9k}
}

@misc{Yang2026b,
  author        = {Yang, Yang and Rayat, Ali and Jang, Ho and Fan, Yunhao and Bakshi, Sankha Subhra and Chern, Gia-Wei},
  title         = {{Adiabatic Variational Principle}: {Born--Oppenheimer} Theory for Emergent Electronic-Order Dynamics},
  year          = {2026},
  eprint        = {2610.40117},
  archivePrefix = {arXiv},
  url           = {https://arxiv.org/abs/2610.40117}
}

@article{Chern2019,
  author        = {Chern, Gia-Wei and Barros, Kipton},
  title         = {Nonequilibrium dynamics of superconductivity in the attractive {Hubbard} model},
  journal       = {Physical Review B},
  volume        = {99},
  pages         = {035162},
  year          = {2019},
  doi           = {10.1103/PhysRevB.99.035162},
  url           = {https://doi.org/10.1103/PhysRevB.99.035162}
}

@article{Bardeen1957,
  author        = {Bardeen, J. and Cooper, L. N. and Schrieffer, J. R.},
  title         = {Theory of Superconductivity},
  journal       = {Physical Review},
  volume        = {108},
  pages         = {1175--1204},
  year          = {1957},
  doi           = {10.1103/PhysRev.108.1175},
  url           = {https://doi.org/10.1103/PhysRev.108.1175}
}

@book{deGennes1966,
  author        = {de Gennes, Pierre-Gilles},
  title         = {Superconductivity of Metals and Alloys},
  publisher     = {W. A. Benjamin},
  year          = {1966},
  address       = {New York}
}

@book{Tinkham1996,
  author        = {Tinkham, Michael},
  title         = {Introduction to Superconductivity},
  publisher     = {McGraw-Hill},
  year          = {1996},
  edition       = {2},
  address       = {New York}
}

@article{Micnas1988,
  author        = {Micnas, R. and Ranninger, J. and Robaszkiewicz, S. and Tabor, S.},
  title         = {Superconductivity in a narrow-band system with intersite electron pairing in two dimensions: A mean-field study},
  journal       = {Physical Review B},
  volume        = {37},
  pages         = {9410--9422},
  year          = {1988},
  doi           = {10.1103/PhysRevB.37.9410},
  url           = {https://doi.org/10.1103/PhysRevB.37.9410}
}

@article{Micnas1990,
  author        = {Micnas, R. and Ranninger, J. and Robaszkiewicz, S.},
  title         = {Superconductivity in narrow-band systems with local nonretarded attractive interactions},
  journal       = {Reviews of Modern Physics},
  volume        = {62},
  pages         = {113},
  year          = {1990},
  doi           = {10.1103/RevModPhys.62.113},
  url           = {https://doi.org/10.1103/RevModPhys.62.113}
}

@article{Chen2023,
  author        = {Chen, Wei-Chih and Wang, Yao and Chen, Cheng-Chien},
  title         = {Superconducting phases of the square-lattice extended {Hubbard} model},
  journal       = {Physical Review B},
  volume        = {108},
  pages         = {064514},
  year          = {2023},
  doi           = {10.1103/PhysRevB.108.064514},
  url           = {https://doi.org/10.1103/PhysRevB.108.064514}
}

@article{Gygi1991,
  author        = {Gygi, Fran{\c{c}}ois and Schl{\"u}ter, Michael},
  title         = {Self-consistent electronic structure of a vortex line in a type-{II} superconductor},
  journal       = {Physical Review B},
  volume        = {43},
  pages         = {7609--7621},
  year          = {1991},
  doi           = {10.1103/PhysRevB.43.7609},
  url           = {https://doi.org/10.1103/PhysRevB.43.7609}
}

@article{Ghosal2000,
  author        = {Ghosal, Amit and Randeria, Mohit and Trivedi, Nandini},
  title         = {Spatial inhomogeneities in disordered {$d$}-wave superconductors},
  journal       = {Physical Review B},
  volume        = {63},
  pages         = {020505},
  year          = {2000},
  doi           = {10.1103/PhysRevB.63.020505},
  url           = {https://doi.org/10.1103/PhysRevB.63.020505},
  note          = {Published online 20 December 2000; appears in volume 63 (2001)}
}

@article{Wang1995,
  author        = {Wang, Yong and MacDonald, A. H.},
  title         = {Mixed-state quasiparticle spectrum for {$d$}-wave superconductors},
  journal       = {Physical Review B},
  volume        = {52},
  pages         = {R3876},
  year          = {1995},
  doi           = {10.1103/PhysRevB.52.R3876},
  url           = {https://doi.org/10.1103/PhysRevB.52.R3876}
}

@article{Franz1998,
  author        = {Franz, M. and Te{\v{s}}anovi{\'c}, Z.},
  title         = {Self-Consistent Electronic Structure of a {$d_{x^2-y^2}$} and a {$d_{x^2-y^2}+id_{xy}$} Vortex},
  journal       = {Physical Review Letters},
  volume        = {80},
  pages         = {4763},
  year          = {1998},
  doi           = {10.1103/PhysRevLett.80.4763},
  url           = {https://doi.org/10.1103/PhysRevLett.80.4763}
}

@article{Berthod2016,
  author        = {Berthod, Christophe},
  title         = {Vortex spectroscopy in the vortex glass: A real-space numerical approach},
  journal       = {Physical Review B},
  volume        = {94},
  pages         = {184510},
  year          = {2016},
  doi           = {10.1103/PhysRevB.94.184510},
  url           = {https://doi.org/10.1103/PhysRevB.94.184510}
}

@article{Sigrist1991,
  author        = {Sigrist, Manfred and Ueda, Kazuo},
  title         = {Phenomenological theory of unconventional superconductivity},
  journal       = {Reviews of Modern Physics},
  volume        = {63},
  pages         = {239},
  year          = {1991},
  doi           = {10.1103/RevModPhys.63.239},
  url           = {https://doi.org/10.1103/RevModPhys.63.239}
}

@article{Tsuei2000,
  author        = {Tsuei, C. C. and Kirtley, J. R.},
  title         = {Pairing symmetry in cuprate superconductors},
  journal       = {Reviews of Modern Physics},
  volume        = {72},
  pages         = {969--1016},
  year          = {2000},
  doi           = {10.1103/RevModPhys.72.969},
  url           = {https://doi.org/10.1103/RevModPhys.72.969}
}

@article{Scalapino2012,
  author        = {Scalapino, D. J.},
  title         = {A common thread: The pairing interaction for unconventional superconductors},
  journal       = {Reviews of Modern Physics},
  volume        = {84},
  pages         = {1383},
  year          = {2012},
  doi           = {10.1103/RevModPhys.84.1383},
  url           = {https://doi.org/10.1103/RevModPhys.84.1383}
}

@article{Pan2000,
  author        = {Pan, S. H. and Hudson, E. W. and Gupta, A. K. and Ng, K.-W. and Eisaki, H. and Uchida, S. and Davis, J. C.},
  title         = {{STM} Studies of the Electronic Structure of Vortex Cores in {$\mathrm{Bi}_2\mathrm{Sr}_2\mathrm{CaCu}_2\mathrm{O}_{8+\delta}$}},
  journal       = {Physical Review Letters},
  volume        = {85},
  pages         = {1536},
  year          = {2000},
  doi           = {10.1103/PhysRevLett.85.1536},
  url           = {https://doi.org/10.1103/PhysRevLett.85.1536}
}

@article{Gorkov1959,
  author        = {Gor'kov, L. P.},
  title         = {Microscopic Derivation of the {Ginzburg--Landau} Equations in the Theory of Superconductivity},
  journal       = {Soviet Physics JETP},
  volume        = {9},
  pages         = {1364--1367},
  year          = {1959}
}

@article{Abrahams1966,
  author        = {Abrahams, Elihu and Tsuneto, Toshihiko},
  title         = {Time Variation of the {Ginzburg--Landau} Order Parameter},
  journal       = {Physical Review},
  volume        = {152},
  pages         = {416},
  year          = {1966},
  doi           = {10.1103/PhysRev.152.416},
  url           = {https://doi.org/10.1103/PhysRev.152.416}
}

@article{Schmid1966,
  author        = {Schmid, Albert},
  title         = {A time dependent {Ginzburg--Landau} equation and its application to the problem of resistivity in the mixed state},
  journal       = {Physik der Kondensierten Materie},
  volume        = {5},
  pages         = {302--317},
  year          = {1966},
  doi           = {10.1007/BF02422669},
  url           = {https://doi.org/10.1007/BF02422669}
}

@article{Kosterlitz1973,
  author        = {Kosterlitz, J. M. and Thouless, D. J.},
  title         = {Ordering, metastability and phase transitions in two-dimensional systems},
  journal       = {Journal of Physics C: Solid State Physics},
  volume        = {6},
  pages         = {1181--1203},
  year          = {1973},
  doi           = {10.1088/0022-3719/6/7/010},
  url           = {https://doi.org/10.1088/0022-3719/6/7/010}
}

@article{Yurke1993,
  author        = {Yurke, B. and Pargellis, A. N. and Kovacs, T. and Huse, D. A.},
  title         = {Coarsening dynamics of the {$XY$} model},
  journal       = {Physical Review E},
  volume        = {47},
  pages         = {1525--1530},
  year          = {1993},
  doi           = {10.1103/PhysRevE.47.1525},
  url           = {https://doi.org/10.1103/PhysRevE.47.1525}
}

@article{Jelic2011,
  author        = {Jeli{\'c}, Asja and Cugliandolo, Leticia F.},
  title         = {Quench dynamics of the 2d {$XY$} model},
  journal       = {Journal of Statistical Mechanics: Theory and Experiment},
  volume        = {2011},
  pages         = {P02032},
  year          = {2011},
  doi           = {10.1088/1742-5468/2011/02/P02032},
  url           = {https://doi.org/10.1088/1742-5468/2011/02/P02032},
  number        = {02}
}

@article{Barankov2004,
  author        = {Barankov, R. A. and Levitov, L. S. and Spivak, B. Z.},
  title         = {Collective {Rabi} Oscillations and Solitons in a Time-Dependent {BCS} Pairing Problem},
  journal       = {Physical Review Letters},
  volume        = {93},
  pages         = {160401},
  year          = {2004},
  doi           = {10.1103/PhysRevLett.93.160401},
  url           = {https://doi.org/10.1103/PhysRevLett.93.160401}
}

@article{Yuzbashyan2006,
  author        = {Yuzbashyan, Emil A. and Dzero, Maxim},
  title         = {Dynamical Vanishing of the Order Parameter in a Fermionic Condensate},
  journal       = {Physical Review Letters},
  volume        = {96},
  pages         = {230404},
  year          = {2006},
  doi           = {10.1103/PhysRevLett.96.230404},
  url           = {https://doi.org/10.1103/PhysRevLett.96.230404}
}

@article{Emery1995,
  author  = {Emery, V. J. and Kivelson, S. A.},
  title   = {Importance of phase fluctuations in superconductors with small superfluid density},
  journal = {Nature},
  volume  = {374},
  pages   = {434--437},
  year    = {1995},
  doi     = {10.1038/374434a0},
  url     = {https://doi.org/10.1038/374434a0}
}

@article{Huang2013,
  author  = {Huang, Wen-Min and Lai, Chen-Yen and Shi, Chuntai and Tsai, Shan-Wen},
  title   = {Unconventional superconducting phases for the two-dimensional extended {Hubbard} model on a square lattice},
  journal = {Phys. Rev. B},
  volume  = {88},
  number  = {5},
  pages   = {054504},
  year    = {2013},
  doi     = {10.1103/PhysRevB.88.054504},
  url     = {https://doi.org/10.1103/PhysRevB.88.054504}
}

@article{Bietenholz2013,
  author  = {Bietenholz, Wolfgang and Gerber, Urs and Rej{\'o}n-Barrera, Fernando G.},
  title   = {{Berezinskii--Kosterlitz--Thouless} transition with a constraint lattice action},
  journal = {J. Stat. Mech.: Theory Exp.},
  volume  = {2013},
  number  = {12},
  pages   = {P12009},
  year    = {2013},
  doi     = {10.1088/1742-5468/2013/12/P12009},
  eprint  = {1307.0485},
  archivePrefix = {arXiv},
  primaryClass = {hep-lat},
  url     = {https://doi.org/10.1088/1742-5468/2013/12/P12009}
}

@article{Barzilai1988,
  author  = {Barzilai, Jonathan and Borwein, Jonathan M.},
  title   = {Two-Point Step Size Gradient Methods},
  journal = {IMA J. Numer. Anal.},
  volume  = {8},
  number  = {1},
  pages   = {141--148},
  year    = {1988},
  doi     = {10.1093/imanum/8.1.141},
  url     = {https://doi.org/10.1093/imanum/8.1.141}
}

@article{Bezanson2017,
  author  = {Bezanson, Jeff and Edelman, Alan and Karpinski, Stefan and Shah, Viral B.},
  title   = {{Julia}: A Fresh Approach to Numerical Computing},
  journal = {SIAM Rev.},
  volume  = {59},
  number  = {1},
  pages   = {65--98},
  year    = {2017},
  doi     = {10.1137/141000671},
  url     = {https://doi.org/10.1137/141000671}
}

@article{Besard2019,
  author  = {Besard, Tim and Foket, Christophe and {De Sutter}, Bjorn},
  title   = {Effective Extensible Programming: Unleashing {Julia} on {GPUs}},
  journal = {IEEE Trans. Parallel Distrib. Syst.},
  volume  = {30},
  number  = {4},
  pages   = {827--841},
  year    = {2019},
  doi     = {10.1109/TPDS.2018.2872064},
  url     = {https://doi.org/10.1109/TPDS.2018.2872064}
}

@article{Barankov2006,
  author        = {Barankov, R. A. and Levitov, L. S.},
  title         = {Synchronization in the {BCS} Pairing Dynamics as a Critical Phenomenon},
  journal       = {Physical Review Letters},
  volume        = {96},
  number        = {23},
  pages         = {230403},
  year          = {2006},
  doi           = {10.1103/PhysRevLett.96.230403},
  url           = {https://doi.org/10.1103/PhysRevLett.96.230403}
}

@article{Foster2013,
  author        = {Foster, Matthew S. and Dzero, Maxim and Gurarie, Victor and Yuzbashyan, Emil A.},
  title         = {Quantum quench in a {$p+ip$} superfluid: Winding numbers and topological states far from equilibrium},
  journal       = {Physical Review B},
  volume        = {88},
  number        = {10},
  pages         = {104511},
  year          = {2013},
  doi           = {10.1103/PhysRevB.88.104511},
  url           = {https://doi.org/10.1103/PhysRevB.88.104511}
}

@article{Foster2014,
  author        = {Foster, Matthew S. and Gurarie, Victor and Dzero, Maxim and Yuzbashyan, Emil A.},
  title         = {Quench-Induced {Floquet} Topological {$p$}-Wave Superfluids},
  journal       = {Physical Review Letters},
  volume        = {113},
  number        = {7},
  pages         = {076403},
  year          = {2014},
  doi           = {10.1103/PhysRevLett.113.076403},
  url           = {https://doi.org/10.1103/PhysRevLett.113.076403}
}

@article{Monaco2002,
  author       = {Monaco, R. and Mygind, J. and Rivers, R. J.},
  title        = {Zurek-Kibble Domain Structures: The Dynamics of Spontaneous Vortex Formation in Annular Josephson Tunnel Junctions},
  journal      = {Physical Review Letters},
  volume       = {89},
  number       = {8},
  pages        = {080603},
  year         = {2002},
  doi          = {10.1103/PhysRevLett.89.080603},
  url          = {https://doi.org/10.1103/PhysRevLett.89.080603}
}

@article{Maniv2003,
  author       = {Maniv, A. and Polturak, E. and Koren, G.},
  title        = {Observation of Magnetic Flux Generated Spontaneously During a Rapid Quench of Superconducting Films},
  journal      = {Physical Review Letters},
  volume       = {91},
  number       = {19},
  pages        = {197001},
  year         = {2003},
  doi          = {10.1103/PhysRevLett.91.197001},
  url          = {https://doi.org/10.1103/PhysRevLett.91.197001}
}

@article{Golubchik2010,
  author       = {Golubchik, Daniel and Polturak, Emil and Koren, Gad},
  title        = {Evidence for Long-Range Correlations within Arrays of Spontaneously Created Magnetic Vortices in a {Nb} Thin-Film Superconductor},
  journal      = {Physical Review Letters},
  volume       = {104},
  number       = {24},
  pages        = {247002},
  year         = {2010},
  doi          = {10.1103/PhysRevLett.104.247002},
  url          = {https://doi.org/10.1103/PhysRevLett.104.247002}
}

@article{Drautz2019,
  author       = {Drautz, Ralf},
  title        = {Atomic cluster expansion for accurate and transferable interatomic potentials},
  journal      = {Physical Review B},
  volume       = {99},
  number       = {1},
  pages        = {014104},
  year         = {2019},
  doi          = {10.1103/PhysRevB.99.014104},
  url          = {https://doi.org/10.1103/PhysRevB.99.014104}
}

@inproceedings{Batatia2022,
  author       = {Batatia, Ilyes and Kov{\'a}cs, D{\'a}vid P{\'e}ter and Simm, Gregor N. C. and Ortner, Christoph and Cs{\'a}nyi, G{\'a}bor},
  title        = {{MACE}: Higher Order Equivariant Message Passing Neural Networks for Fast and Accurate Force Fields},
  booktitle    = {Advances in Neural Information Processing Systems},
  volume       = {35},
  pages        = {11423--11436},
  year         = {2022},
  doi          = {10.52202/068431-0830},
  url          = {https://doi.org/10.52202/068431-0830}
}

@article{Fernandes2014,
  author       = {Fernandes, R. M. and Chubukov, A. V. and Schmalian, J.},
  title        = {What drives nematic order in iron-based superconductors?},
  journal      = {Nature Physics},
  volume       = {10},
  number       = {2},
  pages        = {97--104},
  year         = {2014},
  doi          = {10.1038/nphys2877},
  url          = {https://doi.org/10.1038/nphys2877}
}

@article{Tranquada1995,
  author       = {Tranquada, J. M. and Sternlieb, B. J. and Axe, J. D. and Nakamura, Y. and Uchida, S.},
  title        = {Evidence for stripe correlations of spins and holes in copper oxide superconductors},
  journal      = {Nature},
  volume       = {375},
  number       = {6532},
  pages        = {561--563},
  year         = {1995},
  doi          = {10.1038/375561a0},
  url          = {https://doi.org/10.1038/375561a0}
}

@article{Berg2007,
  author       = {Berg, E. and Fradkin, E. and Kim, E.-A. and Kivelson, S. A. and Oganesyan, V. and Tranquada, J. M. and Zhang, S. C.},
  title        = {Dynamical Layer Decoupling in a Stripe-Ordered High-{$T_c$} Superconductor},
  journal      = {Physical Review Letters},
  volume       = {99},
  number       = {12},
  pages        = {127003},
  year         = {2007},
  doi          = {10.1103/PhysRevLett.99.127003},
  url          = {https://doi.org/10.1103/PhysRevLett.99.127003}
}

@article{Hamidian2016,
  author       = {Hamidian, M. H. and Edkins, S. D. and Joo, Sang Hyun and Kostin, A. and Eisaki, H. and Uchida, S. and Lawler, M. J. and Kim, E.-A. and Mackenzie, A. P. and Fujita, K. and Lee, Jinho and Davis, J. C. S{\'e}amus},
  title        = {Detection of a {Cooper}-pair density wave in {Bi$_2$Sr$_2$CaCu$_2$O$_{8+x}$}},
  journal      = {Nature},
  volume       = {532},
  number       = {7599},
  pages        = {343--347},
  year         = {2016},
  doi          = {10.1038/nature17411},
  url          = {https://doi.org/10.1038/nature17411}
}

@article{Edkins2019,
  author       = {Edkins, S. D. and Kostin, A. and Fujita, K. and Mackenzie, A. P. and Eisaki, H. and Uchida, S. and Sachdev, Subir and Lawler, Michael J. and Kim, E.-A. and Davis, J. C. S{\'e}amus and Hamidian, M. H.},
  title        = {Magnetic field--induced pair density wave state in the cuprate vortex halo},
  journal      = {Science},
  volume       = {364},
  number       = {6444},
  pages        = {976--980},
  year         = {2019},
  doi          = {10.1126/science.aat1773},
  url          = {https://doi.org/10.1126/science.aat1773}
}

@article{Dagotto1994b,
  author       = {Dagotto, E. and Riera, J. and Chen, Y. C. and Moreo, A. and Nazarenko, A. and Alcaraz, F. and Ortolani, F.},
  title        = {Superconductivity near phase separation in models of correlated electrons},
  journal      = {Physical Review B},
  volume       = {49},
  number       = {5},
  pages        = {3548--3565},
  year         = {1994},
  doi          = {10.1103/PhysRevB.49.3548},
  url          = {https://doi.org/10.1103/PhysRevB.49.3548}
}

@article{Bishop2016,
  author       = {Bishop, Christopher B. and Liu, Guangkun and Dagotto, Elbio and Moreo, Adriana},
  title        = {On-site attractive multiorbital Hamiltonian for {$d$}-wave superconductors},
  journal      = {Physical Review B},
  volume       = {93},
  number       = {22},
  pages        = {224519},
  year         = {2016},
  doi          = {10.1103/PhysRevB.93.224519},
  url          = {https://doi.org/10.1103/PhysRevB.93.224519}
}

@article{Fisher1991,
  author       = {Fisher, Daniel S. and Fisher, Matthew P. A. and Huse, David A.},
  title        = {Thermal fluctuations, quenched disorder, phase transitions, and transport in type-II superconductors},
  journal      = {Physical Review B},
  volume       = {43},
  number       = {1},
  pages        = {130--159},
  year         = {1991},
  doi          = {10.1103/PhysRevB.43.130},
  url          = {https://doi.org/10.1103/PhysRevB.43.130}
}

@article{Mross2015,
  author       = {Mross, David F. and Senthil, T.},
  title        = {Spin- and Pair-Density-Wave Glasses},
  journal      = {Physical Review X},
  volume       = {5},
  number       = {3},
  pages        = {031008},
  year         = {2015},
  doi          = {10.1103/PhysRevX.5.031008},
  url          = {https://doi.org/10.1103/PhysRevX.5.031008}
}

@article{Reichhardt2020,
  author       = {Reichhardt, Charles and Reichhardt, Cynthia J. O.},
  title        = {Jamming, fragility and pinning phenomena in superconducting vortex systems},
  journal      = {Scientific Reports},
  volume       = {10},
  number       = {1},
  pages        = {11625},
  year         = {2020},
  doi          = {10.1038/s41598-020-68417-0},
  url          = {https://doi.org/10.1038/s41598-020-68417-0}
}

\end{document}